\documentclass[usenatbib,a4paper]{mn2e}

\usepackage{graphicx}
\usepackage{natbib}
\newif\ifAMStwofonts
\usepackage{dcolumn}
\usepackage{bm}
\usepackage{amsmath}    
\usepackage{amsfonts}
\usepackage{amssymb}
\newcommand{\ra}{\;\raise1.0pt\hbox{$'$}\hskip-6pt\partial\;}
\newcommand{\lo}{\;\overline{\raise1.0pt\hbox{$'$}\hskip-6pt\partial}\;}

\newcommand{\etal}{{\em et al.}}

\def\lm{{lm}}
\def\elm{{E,lm}}
\def\blm{{B,lm}}
\def\Yo{{Y}}

\def\beqra{\begin{eqnarray}}
\def\eeqra{\end{eqnarray}}
\def\beq{\begin{equation}}
\def\eeq{\end{equation}}

\def\fr{\frac}

\def \prrd {Phys.\ Rev.\ D\ }

\def\apj{Astrophys. J.\ }

\def\apjl{Astrophys. J. Lett.\ }
\def\apjs{Astrophys. J. Suppl.\ }
\def\mnras{Mon. Not. R. Astron. Soc.\ }

\def\physrep{Phys. Rept.\ }

\newcommand{\bi}{\begin{itemize}}
\newcommand{\ei}{\end{itemize}}
\newcommand{\be}{\begin{equation}}
\newcommand{\ee}{\end{equation}}
\newcommand{\bea}{\begin{eqnarray}}
\newcommand{\eea}{\end{eqnarray}}

\newcommand{\bfn}{\hat{\bf n}}

\newcommand{ \ie }{{\it i.e.}}

\def\spose#1{\hbox to 0pt{#1\hss}}
\def\approxgt{\mathrel{\spose{\lower 3pt\hbox{$\sim$}}
        \raise 2.0pt\hbox{$>$}}}

\renewcommand{\textsf}[1]{{\small #1}}

\newcommand\eq{\begin{equation}}
\newcommand\en{\end{equation}}

\def\edth{\;\raise1.0pt\hbox{$'$}\hskip-6pt\partial\;}
\def\baredth{\;\overline{\raise1.0pt\hbox{$'$}\hskip-6pt
\partial}\;}
\def\bi#1{\hbox{\boldmath{$#1$}}}
\def\gsim{\raise2.90pt\hbox{$\scriptstyle
>$} \hspace{-6.4pt}
\lower.5pt\hbox{$\scriptscriptstyle
\sim$}\; }
\def\lsim{\raise2.90pt\hbox{$\scriptstyle
<$} \hspace{-6pt}\lower.5pt\hbox{$\scriptscriptstyle\sim$}\; }

\title[Lensed CMB maps from the Millennium Simulation]{Lensed CMB temperature and polarization maps from the Millennium Simulation}
\author[]{%
Carmelita Carbone$^{1}$\footnotemark[1],
Carlo Baccigalupi$^{2}$\footnotemark[3],
Matthias Bartelmann$^{3}$\footnotemark[4],
\newauthor
Sabino Matarrese$^{4}$\footnotemark[5],
Volker Springel$^{5}$\footnotemark[2]
\\
\\ 
$^{1}$ Institut de Ci\`encies de l'Espai, CSIC/IEEC, Campus UAB, F. de Ci\`encies, Torre C5 par-2,  Barcelona 08193, Spain \\
$^{2}$ SISSA/ISAS, Astrophysics Sector, Via Beirut 4, I-34014, Trieste, Italy and \\
INFN, Sezione di Trieste, Via Valerio, 2, 34127, Trieste, Italy \\
$^{3}$ Institut f$\ddot{\rm u}$r Theoretische Astrophysik,
Universit$\ddot{\rm a}$t Heidelberg, Tiergartenstrasse 15, D-69121,
Heidelberg, Germany \\
$^{4}$ Dipartimento di Fisica `Galileo Galilei', Universit\`a di Padova and \\ 
INFN, Sezione di Padova, Via Marzolo 8, I-35131 Padova, Italy \\
$^{5}$ Max-Planck-Institute for Astrophysics, Karl-Schwarzschild-Str. 1, D-85741
Garching, Germany\\
}

\begin{document}

\maketitle

\begin{abstract}
We have constructed the first all-sky CMB temperature and polarization
lensed maps based on a high-resolution cosmological $N$-body
simulation, the Millennium Simulation (MS).  We have exploited the
lensing potential map obtained using a map-making procedure
\citep{Carbone_etal2008} which integrates along the
line-of-sight the MS dark matter distribution by stacking and
randomizing the simulation boxes up to $z=127$, and which
semi-analytically supplies the large-scale power in the angular
lensing potential that is not correctly sampled by the $N$-body
simulation. The lensed sky has been obtained by properly modifying the
latest version of the LensPix code \citep{Lewis05} 
to account for the MS structures.  We
have also produced all-sky lensed maps of the so-called $\psi_E$ and
$\psi_B$ potentials, which are directly related to the \emph{electric}
and \emph{magnetic} types of polarization.  The angular power spectra
of the simulated lensed temperature and polarization maps agree well
with semi-analytic estimates up to $l \leq 2500$, while on smaller
scales we find a slight excess of power which we interpret as being
due to non-linear clustering in the MS.  We also observe how
non-linear lensing power in the polarised CMB is transferred to large
angular scales by suitably misaligned modes in the CMB and the lensing
potential. This work is relevant in view of the future CMB probes, as
a way to analyse the lensed sky and disentangle the contribution from
primordial gravitational waves.
\end{abstract}


\section{Introduction}
\renewcommand{\thefootnote}{\fnsymbol{footnote}}
\footnotetext[1]{E-mail: carbone@ieec.uab.es}
\footnotetext[5]{E-mail: bacci@sissa.it}
\footnotetext[3]{E-mail: mbartelmann@ita.uni-heidelberg.de}
\footnotetext[4]{E-mail: sabino.matarrese@pd.infn.it}
\footnotetext[2]{E-mail: volker@MPA-Garching.MPG.DE}
\renewcommand{\thefootnote}{\arabic{footnote}}

The Cosmic Microwave Background (CMB) is characterized both by
\emph{primary} anisotropies, imprinted at the last scattering surface
at redshift $z\sim 1100$, and by \emph{secondary} anisotropies caused
along the way to us by density inhomogeneities and re-scattering off
electrons that are freed during the epoch of reionization, and heated
to high temperature when massive structures virialize.

On one hand, the primary CMB anisotropies give direct insight into the
structure of the very early Universe, and are one of the principal
pillars on which the standard cosmological $\Lambda$CDM model is
founded.  The temperature anisotropy power spectrum has now been
measured to very high precision \citep[e.g.][]{WMAP5} yielding tight
constraints on the basic parameters of the cosmological model.

On the other hand, one of the most important mechanisms that can
generate secondary anisotropies is the weak gravitational lensing of
the CMB, which arises from the distortions induced in the geodesics of
CMB photons by gradients in the gravitational matter potential
\citep{Matthias_rew,Lewis06}. The remapping of points produced by
lensing induces non-Gaussianities in the observed CMB sky, and also
changes the power spectra of the perturbations.

The CMB is also expected to be polarized at the $\sim10$\% level
principally because of Thomson scattering of photons off free
electrons during recombination.  Thomson scattering generates linear
polarization only, which is usually expressed in terms of the Stokes
parameters $Q$ and $U$.  They can in turn be decomposed into
coordinate independent $E$- and $B$-modes of polarization
\citep[e.g.][]{Zaldarriaga&Seljak97} with opposite parities, the
so-called ``electric'' and ``magnetic'' types of polarization.  To
linear order in perturbation theory, primordial scalar (density)
perturbations can only generate $E$-polarization, while primordial
vector and tensor (gravitational waves) perturbations can generate
both scalar $E$- and pseudoscalar $B$-polarization
\citep[e.g.]{Seljak&Zaldarriaga97}.  In particular, \emph{primary}
B-mode polarization represents the imprints left from the primordial
gravitational waves (GWs) on the CMB \citep{Hu et al. 1998, review1,
  review2}: if initial fluctuations are created very early,
e.g.~during inflation so that the vector growth is damped, primary
B-modes are produced only by tensor perturbations that, being damped
at last scattering by the horizon entering, produce the largest amount
of temperature quadrupole anisotropy and, consequently, by Thomson
scattering, the largest amount of polarization
\citep{Pritchard&Kamionkowski04}.  The primordial gravitational
radiation is thought to have been generated by quantum fluctuations of
the metric tensor during the inflationary era, with a strain amplitude
proportional to the square of the inflation energy scale.
Consequently, the indirect detection of this relic gravitational
background via the direct observation of the primary B-mode
polarization, as expected from future dedicated CMB missions by ESA
and NASA\footnote{See \textsf{lambda.gsfc.nasa.gov} for a complete
  list of operating and planned CMB experiments}, will shed light on
the physics of the very early Universe and will represent a powerful
magnifier on the inflationary era and the very first moments in the
existence of the Universe \citep{Kamionkowski et al. 1997,
  Zaldarriaga&Seljak97, Seljak&Zaldarriaga97}.

However, as for the CMB temperature, there are mechanisms also for the
polarization that can produce \emph{secondary} B-modes, with the
dominant one being again gravitational lensing, i.e. \emph{cosmic
  shear} (CS), which distorts the primary CMB pattern, in particular
converting E- into B-polarization \citep{Zaldarriaga&Seljak98}, even
in case of absence of primary B-modes.  Although comparable, B-modes
from primordial GWs exhibit their peak at multipoles $l\approx 100$,
corresponding to the degree scale, while, for lensed B-modes, the peak
is at $l\approx1000$, corresponding to the arcminute scale.
Nonetheless, if the energy scale of inflation is $V^{1/4}\le
4\times10^{15}$GeV, the CS-induced curl represents a foreground for
the $l\approx50-100$ primordial GW-induced primary B-polarization
\citep{Cabella_Kamionkowski04}.  This could limit the extraction of
the gravity wave signal if not taken into account correctly
\citep[e.g.][]{seljak_hirata_2004}, even though forthcoming CMB probes
will have in principle the sensitivity and the instrumental
performance for the detection of the CMB anisotropies in total
intensity and polarization.

A precise knowledge of the lensing effects would also provide new
insights and constraints on the expansion history of the Universe, on
the process of cosmological structure formation
\citep{acquaviva_baccigalupi_2006,hu_etal_2006} and on the
cosmological parameter estimation \citep{Smith_etal2006,
  Smith_etal2006_bis}.  In particular, for a correct interpretation of
the data from the forthcoming Planck
satellite\footnote{www.rssd.esa.int/PLANCK}, it will be absolutely
essential to understand and model the CMB lensing, as the satellite
has the sensitivity for measuring the CMB lensing with good
accuracy. We note that a first detection of CMB lensing in data from
the Wilkinson Microwave Anisotropy Probe (WMAP\footnote{See
  \textsf{map.gsfc.nasa.gov}}) combined with complementary data has
already been claimed by \cite{smith_etal_2007} and \cite{Seljak0108},
and evidence for weak gravitational lensing of the CMB has been observed
at $>$3-sigma significance by \cite{ACBAR}.

From the arguments above it follows that the next generation CMB
experiments will require a detailed lensing reconstruction and an
accurate de-lensing methodology.  One can try to reconstruct the
gravitational lensing effects using the so-called quadratic and
maximum-likelihood estimators \citep{Hu&Okamoto2002,
  Hirata&Seljak2003}, which allow to reproduce at some level of
precision the lensing potential from the observed CMB itself, and to
invert the photon geodesic remapping induced by the cosmic shear.  Up
to now these methodologies have been applied only to limited patches
of sky \citep{Amblard et al 04} and/or under the hypothesis of a
Gaussian distribution of the lensing sources
\citep[e.g.][]{seljak_hirata_2004}.  However, this is only a first
order approximation, since the non-linear evolution of the cosmic
structures induces non-Gaussian features in the lensing potential,
features that in turn have an impact on the non-Gaussian statistic of
the CMB produced by the point remapping caused by the cosmic shear
itself on the CMB fluctuation pattern.

It is thus very interesting to test the performance of these
estimators on full-sky, lensed CMB maps which do include the effects
of the non-linear structure evolution at all the orders. This demands
detailed simulated lensed CMB maps.

The increasing availability of high-resolution $N$-body simulations in
large periodic volumes makes it possible to directly simulate the CMB
distortions caused by weak lensing using realistic cosmological
structure formation calculations. Our previous work
\citep{Carbone_etal2008} represents a first step into this direction.
Indeed, existing studies already give access to statistical properties
of the expected all-sky CMB lensing signal \citep[see e.g.][and
  references therein]{Lewis05}, but these studies are based on
`semi-analytic' calculations that use approximate parameterizations of
the non-linear evolution of the matter power spectrum.  On the other
hand, up to now $N$-body numerical simulations have been used to lens
the CMB only in limited patches of sky \citep{Das&Bode08}, or to
produce full-sky convergence maps confined to low redshifts $z\sim 1$
\citep{Fosalba07, arXiv:0807.3651}.

In \cite{Carbone_etal2008} we have developed and described a procedure
which gives access to the full statistics of the lensed CMB signal,
including non-linear and non-Gaussian effects on the full-sky. This
should allow improvements in the methods for separating the different
contributions to CMB anisotropies in the data, which would help
substantially to uncover all the cosmological information in the
forthcoming observations.

In this paper we apply the methodology developed in
\cite{Carbone_etal2008} to the construction of all-sky, lensed
simulated temperature and polarization maps.  In Section~2 we describe
the procedure utilized to lens, which is based on the simulated
lensing potential and deflection angle templates obtained in
\citet{Carbone_etal2008}.  In Section~3 we describe the resulting
simulated lensed maps.  In Section~4 and Section~5 we apply
statistical analyses to the obtained temperature and polarization
lensed maps showing consistency and possible differences with respect
to semi-analytical expectations. Finally in Section~6 we draw the
conclusions of this work and outline next steps and future
applications.
\begin{figure}
\includegraphics[width=.48\textwidth]{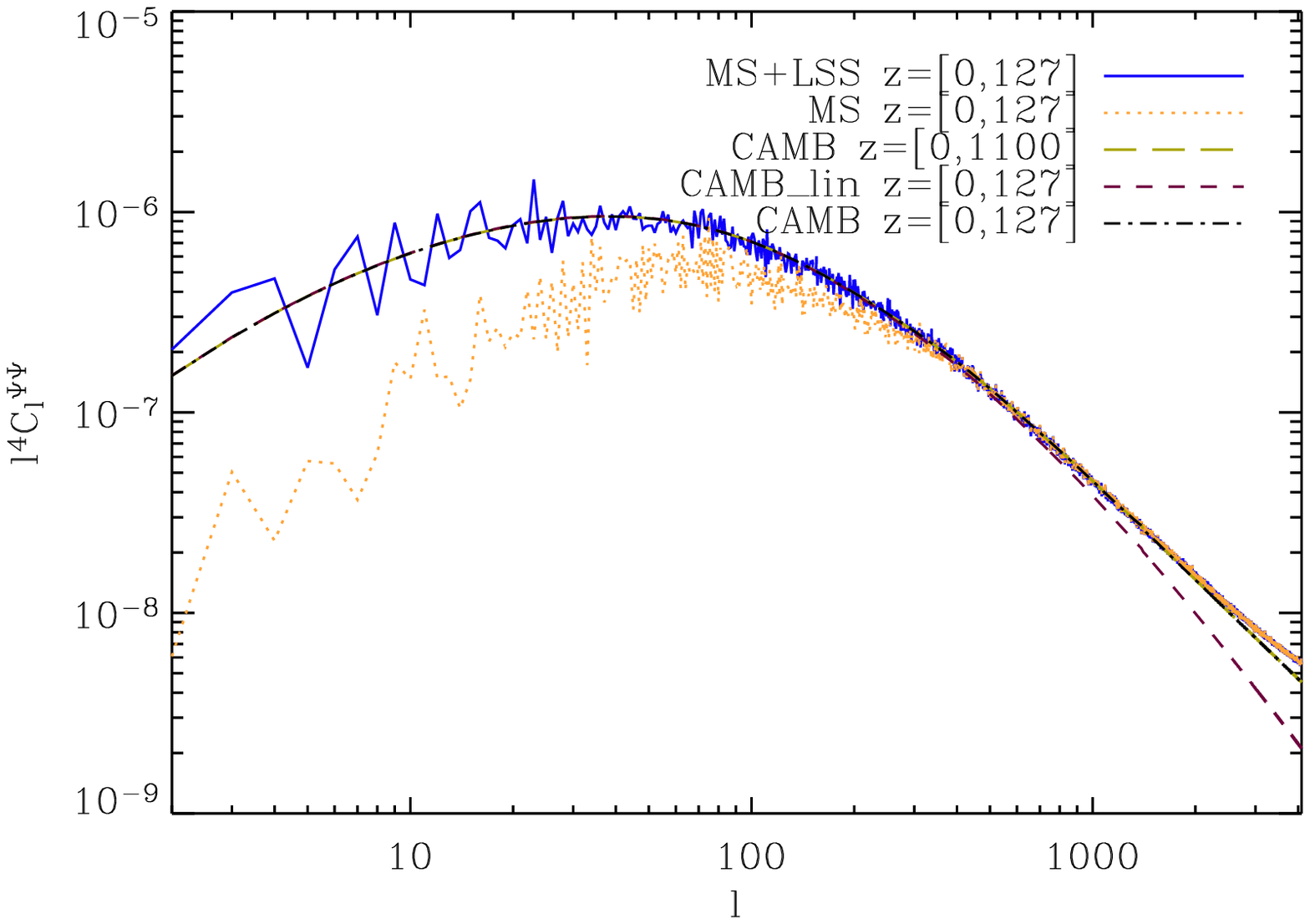}
\includegraphics[width=.48\textwidth]{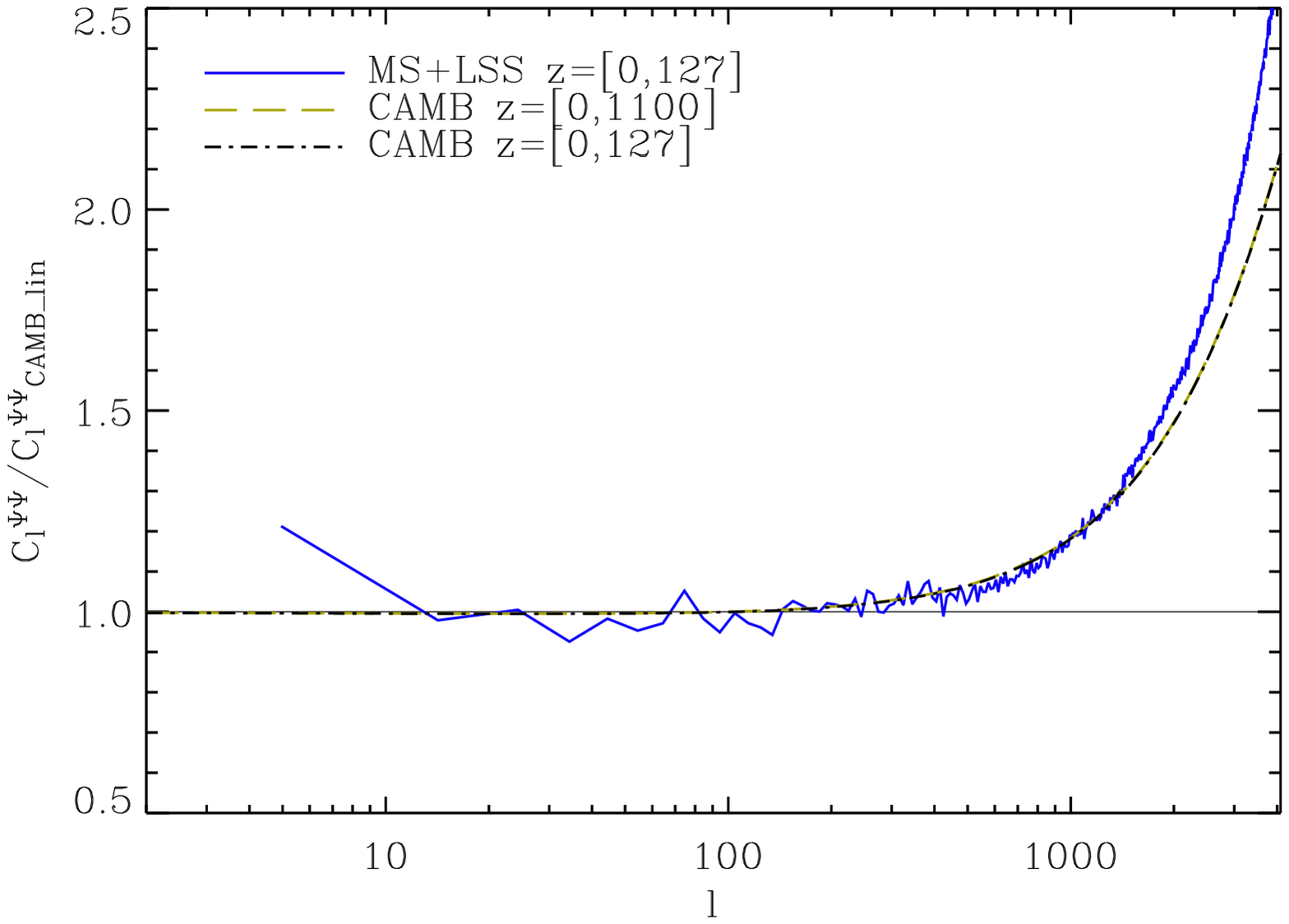}
\footnotesize
\caption{{\em Top panel:} The dotted orange line represents the simulated lensing
  potential power spectrum obtained via line-of-sight integration
  across the MS dark matter distribution up to $z=127$. The blue solid
  line is the same as the orange dotted one after reinstating the
  large scale power with the use of the LS-adding technique. The
  dot-dashed black line represents the power spectrum of the lensing
  potential obtained with the CAMB code stopping the line-of sight
  integration at $z=127$, and including an estimate of the
  non-linear contributions \citep{Halofit}.  The long-dashed light
  green line is the same as the dot-dashed black one with
  line-of-sight integration up to $z=1100$.  Finally, the dashed
  violet line represents the linear lensing potential power spectrum
  from the CAMB code in the linear approximation and integrating up to
  $z=127$.  {\em Bottom panel:} The ratios between the power spectra shown in the
  top panel and the CAMB linear lensing potential spectrum up to
  $z=127$. It is worth to note that there is no difference using
  $z=1100$ or $z=127$ for the semi-analytical expectations.}
\label{projpot_PS}
\end{figure}

\section{CMB lensing through the Millennium Simulation}
Weak lensing of the CMB deflects photons coming from an original
direction ${\bf \hat{n}}'$ on the last scattering surface to a
direction ${\bf \hat{n}}$ on the observed sky, so a lensed CMB field
is given by $\tilde{X}({\bf \hat{n}}) = X({\bf \hat{n}}')$ in terms of
the unlensed field $X=T, Q, U$ \citep[e.g.][]{Lewis05}.  The
displacement of the points is determined by the integral of the
gravitational potential along the line of sight to the last scattering
surface, as we review below.

In what follows we will consider only the \emph{small-angle
  scattering} limit, \ie~the case where the \emph{change} in the
comoving separation of CMB light-rays, owing to the deflection caused
by gravitational lensing from matter inhomogeneities, is small
compared to the comoving separation between the \emph{undeflected}
rays. In this case it is sufficient to calculate all the relevant
integrated quantities, \ie~the so-called \emph{lensing-potential} and
its angular gradient, the \emph{deflection-angle}, along the
undeflected rays. This small-angle scattering limit corresponds to the
Born approximation.

Adopting conformal time and comoving coordinates in a flat geometry
\citep{maber}, the integral for the projected lensing-potential due to
scalar perturbations with no anisotropic stress reads
\begin{align}
\label{lensingpotential}
\Psi({\bf \hat{n}})\equiv
-2\int_0^{r_*} \fr{r_*-r}{r_*r}\,\fr{\Phi(r{\bf\hat{n}};\eta_0-r )}{c^2}\,{\rm d}r\,,
\end{align}
where $r$ is the comoving distance, $r_*\simeq 10^4$ Mpc is its value
at the last-scattering surface, $\eta_0$ is the present conformal
time, and $\Phi$ is the physical peculiar gravitational potential
generated by density perturbations
\citep{Hu2000,Matthias_rew,Refregier,Lewis06}.

Actually, the lensing potential is formally divergent owing to the
$1/r$ term near $r=0$; nonetheless, this divergence affects the
lensing potential monopole only, which can be set to zero, since it
does not contribute to the deflection-angle. In this way the remaining
multipoles take a finite value and the lensing potential field is well
defined \citep{Lewis06}.
\begin{figure*}
\begin{center}
\resizebox{12.7cm}{!}{\includegraphics{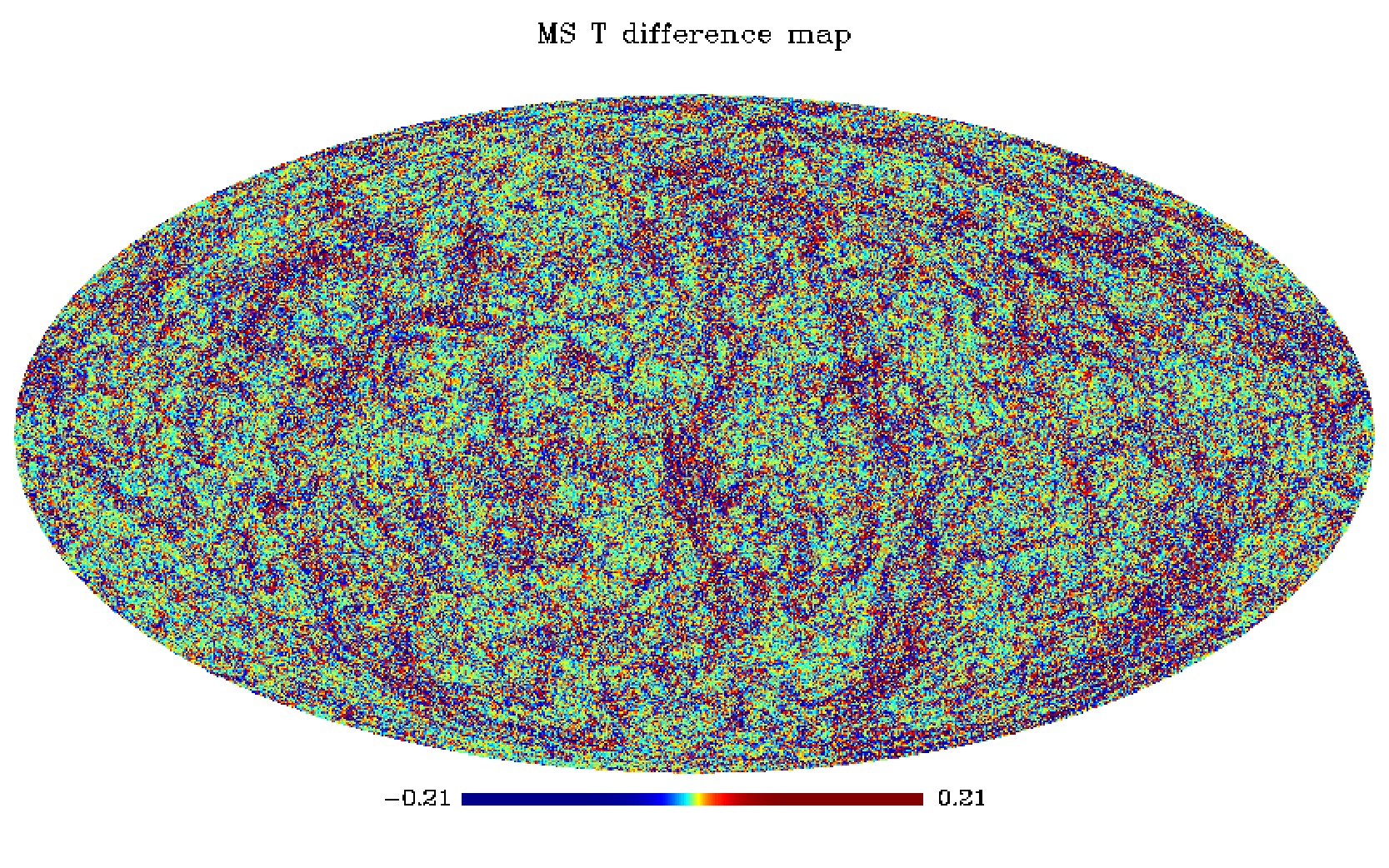}}
\resizebox{12.7cm}{!}{\includegraphics{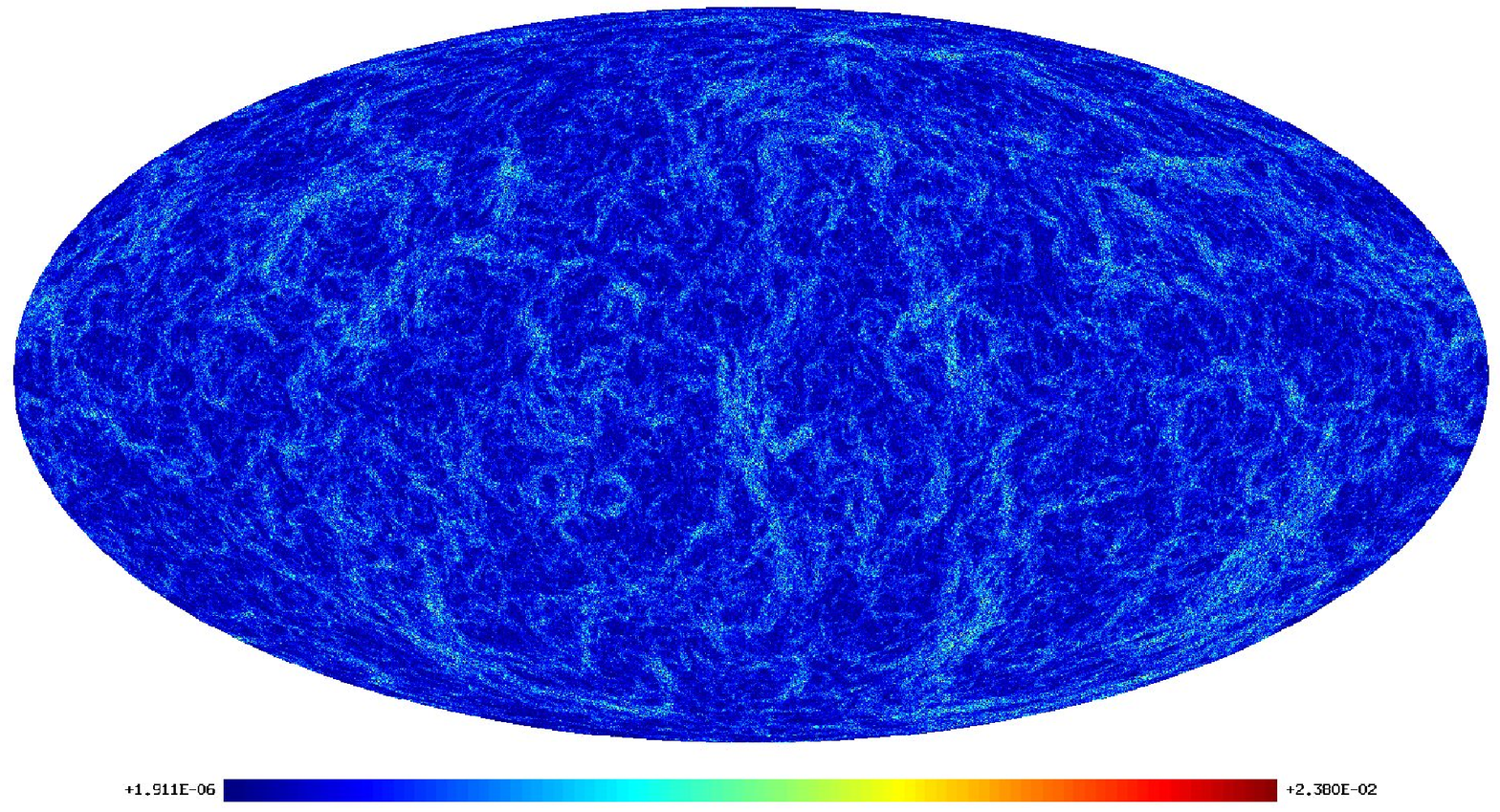}}
\resizebox{12.7cm}{!}{\includegraphics{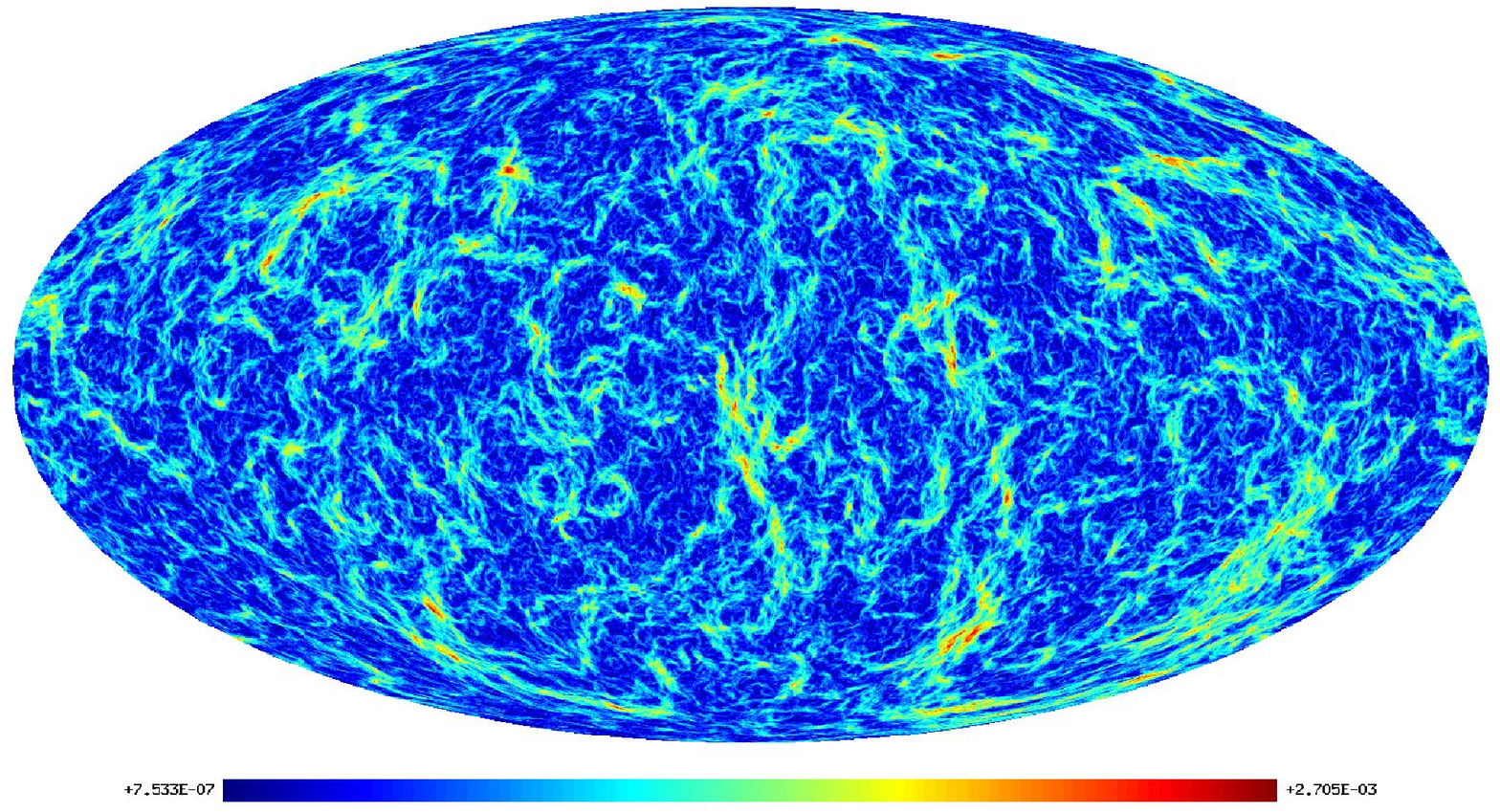}}
\footnotesize
\caption{{\em Top panel:} Difference $T$ map between the lensed and unlensed $T$
  fields obtained by supplying the LensPix code with the spherical
  harmonic coefficients extracted from the MS lensing potential map
  and by implementing in it the LS-adding technique, as described in
  the text. Units in $mK$. A histogram equalized color mapping has
  been used to increase the contrast.  {\em Middle panel:} Modulus of the
  polarization $\Delta P\equiv \sqrt{\Delta Q^2+\Delta U^2}$, where
  $\Delta Q$ and $\Delta U$ are the difference $Q$ and $U$ maps
  obtained using the same technique as for the $T$ difference map of
  the top panel. Units in $mK$.  {\em Bottom panel:} Map of the deflection-angle
  modulus obtained as angular gradient of the lensing-potential whose
  power spectrum is represented by the solid blue line in
  Fig.~\ref{projpot_PS}}
\label{TP_diff_maps}
\end{center}
\end{figure*}
The vector ${\bf \hat{n}}'$ is obtained from ${\bf \hat{n}}$ by moving
its end on the surface of a unit sphere by a distance
$|\nabla_{\bf\hat{n}}\Psi({\bf \hat{n}})|$ along a geodesic in the
direction of $\nabla_{\bf\hat{n}}\Psi({\bf \hat{n}})$, where
$[1/r]\nabla_{\hat{\bf n}}$ is the two dimensional (2D) transverse
derivative with respect to the line-of-sight pointing in the direction
${\hat{\bf n}}\equiv(\vartheta,\varphi)$
\citep{Hu2000,Challinor02,Lewis05}.  We assume
$|\nabla_{\bf\hat{n}}\Psi({\bf \hat{n}})|$ to be constant between
${\bf \hat{n}}$ and ${\bf \hat{n}}'$, consistent with the Born
approximation.

If the gravitational potential $\Phi$ is Gaussian, so is the lensing
potential.  However, the lensed CMB is non-Gaussian, as it is a second
order cosmological effect produced by cosmological perturbations onto
CMB anisotropies, yielding a finite correlation between different
scales and thus non-Gaussianity.  This is expected to be most
important on small scales, due to the non-linearity already present in
the underlying properties of lenses.

In order to generate full-sky $T$, $Q$, $U$ maps lensed by the matter
distribution of the Millennium Simulation, we have modified the
publicly available LensPix
code\footnote{http://cosmologist.info/lenspix/} (LP), which is
described in \cite{Lewis05}. In its original version, in fact, this
code lenses the primary CMB intensity and polarization fields via a
Gaussian realization, in the spherical harmonic domain, of the lensing
potential power spectrum as extracted from the publicly available Code
for Anisotropies in the Microwave Background
(CAMB\footnote{http://camb.info/}).  Our modification (hereafter
referred to as ``MS-modified-LP'') consists in forcing LP to deflect
the CMB photons using the fully non-linear and non-Gaussian lensing
potential realization obtained from the MS using the procedure
briefly summarized here, which was presented by
\cite{Carbone_etal2008}; we refer the reader to that paper for more
detail.

The MS is a high-resolution $N$-body simulation for a $\Lambda$CDM
cosmology consistent with the WMAP first year results
\citep{spergel2003}, carried out by the Virgo Consortium
\citep{Springel2005}. It uses about 10 billion collisionless particles
with mass $8.6\times 10^{8} h^{-1}{\rm M_\odot}$, in a cubic region
$500\,h^{-1}{\rm Mpc}$ on a side which evolves from redshift
$z_{*}=127$ to the present, with periodic boundary conditions.  Our
map-making procedure is based on ray-tracing of the CMB photons in the
Born approximation through the three-dimensional field of the MS
peculiar gravitational potential. In order to produce mock lensing
potential maps that cover the past light-cone over the full sky, we
stack the peculiar gravitational potential grids around the observer
(which is located at $z=0$), exploiting the pre-computed and stored
snapshots of the simulation. The spacing of the time outputs of the MS
simulation is such that it corresponds to an average distance of
$140\,h^{-1}{\rm Mpc}$ (comoving) on the past light-cone. We fully exploit this time resolution
which, at high accuracy, allows to avoid the adoption of time interpolation techniques, 
and use all the 63 outputs of the simulation along
our integration paths. In practice this means that the data
corresponding to a particular output time is utilized in a spherical
shell of average thickness $140\,h^{-1}{\rm Mpc}$.
Moreover, the total volume around the observer up
to $z_{*}$ is divided into spherical shells, each of thickness
$500h^{-1}{\rm Mpc}$. All the MS boxes falling into the same shell are
translated and rotated with the same random vectors generating a
homogeneous coordinate transformation throughout the shell, while
the randomization vectors change from shell to shell. The peculiar gravitational
potential at each point along a ray in direction $\bfn$ is
spatially interpolated from the pre-computed MS grid which possesses a spatial
resolution of about $195h^{-1}{\rm kpc}$. The deflection angle is
computed along the line of sight as well, by numerically evaluating
the gravitational potential gradient and interpolating at each point
along the line of sight \citep{Carbone_etal2008}. 

Being repeated on scales larger than the box size, the resulting weak
lensing distortion lacks large scale power, which manifests itself in
the lensing potential power spectrum as an evident loss of large scale
power with respect to semi-analytic expectations, which is most
noticeable for multipoles smaller than $l\simeq 400$. This has been
cured by augmenting large scale power (LS-adding) directly in the
angular domain, a procedure which we exploit here as well.  More
specifically, we have implemented the LS-adding technique directly
into the LensPix code as we now explain. We have again split
the spherical harmonics domain into two multipole ranges: $0\leq l
\leq 400$, where the MS fails in reproducing the correct lensing
potential power due to the limited box-size of the simulation, and $l
> 400$, where instead the power spectrum is reproduced correctly by
the Millennium Simulation (see Fig.~\ref{projpot_PS}).  On the latter
interval of multipoles, we have extracted the corresponding ensemble
$\Psi_{lm}^{\rm MS}$ of lensing-potential spherical harmonic
coefficients produced by the MS lens distribution.  We have modified
the LensPix code so that it reads and uses these MS harmonic
coefficients on the corresponding range of multipoles.  On the
interval $0\leq l \leq 400$, instead, we let LensPix generate its own
ensemble of spherical harmonic coefficients $\Psi_{lm}^{\rm LP}$,
which are a realization of a Gaussian random field characterised by
the CAMB semi-analytic lensing-potential power spectrum
\citep[including the estimate of the contribution from
  non-linearity][]{Halofit} inserted as input in the parameter file of
LP.

Since on low multipoles the effects of the non-Gaussianity from the
non-linear scales are negligible and the $\Psi_{lm}$ are independent,
every time that we run the MS-modified-LP, we generate a joined
ensemble of $\tilde{\Psi}_{lm}$, where
$\tilde{\Psi}_{lm}=\Psi_{lm}^{\rm LP}$ for $0\leq l \leq 400$ and
$\tilde{\Psi}_{lm}=\Psi_{lm}^{\rm MS}$ for $l > 400$.

This technique achieves two goals: firstly it reproduces correctly the
non-linear and non-Gaussian effects of the MS non-linear dark matter
distribution on multipoles $l > 400$, including at the same time the
contribution from the large scales at $l \le 400$, where the lensing
potential follows mostly the linear trend as shown from the
light-green dot-dashed line in Fig.~\ref{projpot_PS}. Secondly, it
allows to take correctly into account the cross-correlation between
the temperature and the lensing-potential \citep[e.g.][]{Lewis06} due
to the Integrated Sachs-Wolfe (ISW) effect on the low multipoles,
i.e. on the large scales \citep[this effect is instead negligible at
  $l \gtrsim 200$][]{Afshordi04}.  In fact, as it happens for the
semi-analytic contribution to the lensing-potential, also the primary
temperature and polarization fields are generated by LP as Gaussian
realizations, in the harmonic domain, of the corresponding power
spectra obtained by CAMB, which correctly includes the contribution to
the temperature from the ISW effect.
\begin{figure*}
\begin{center}
\resizebox{13cm}{!}{\includegraphics{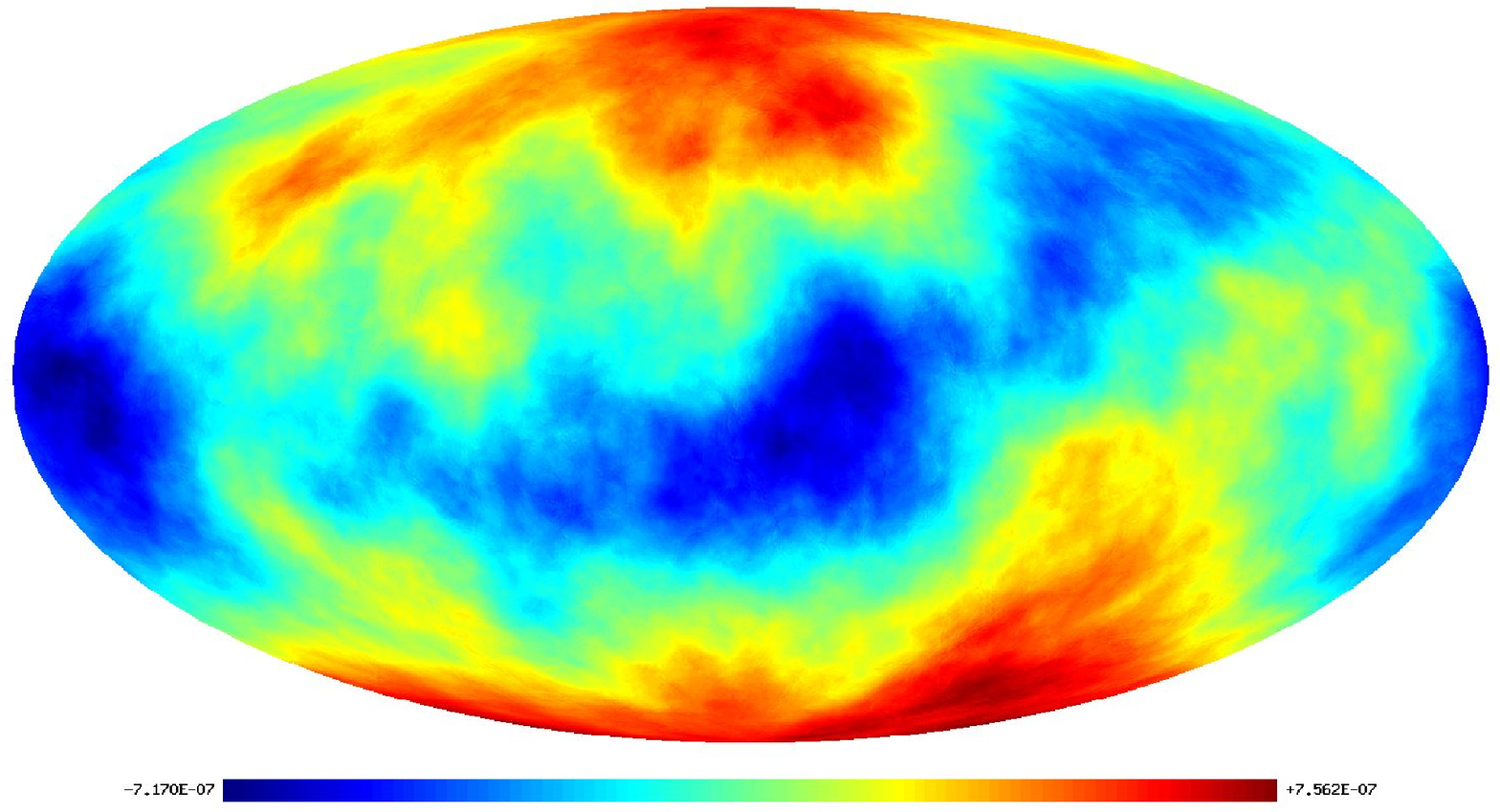}}
\resizebox{13cm}{!}{\includegraphics{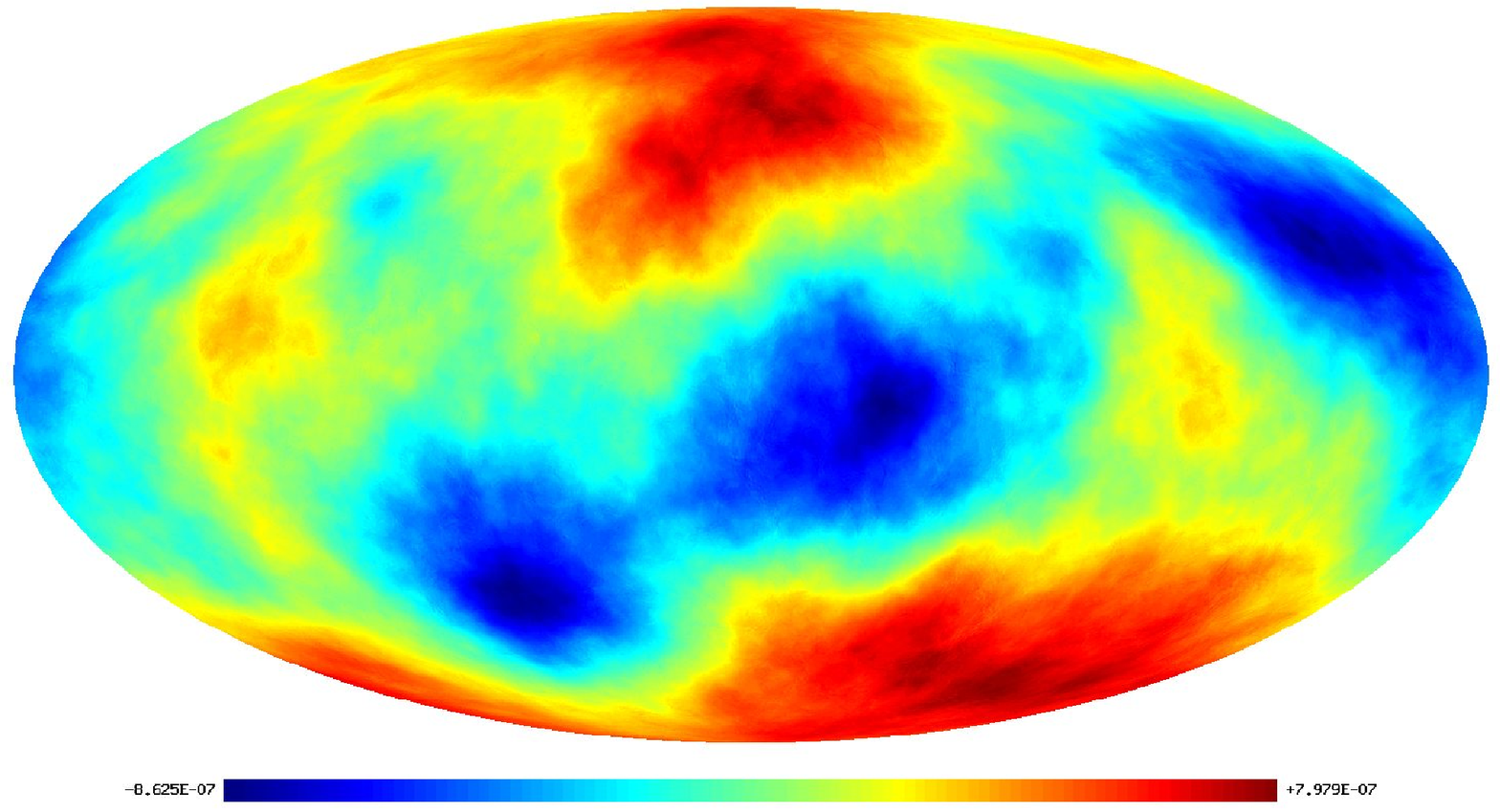}}
\footnotesize
\caption{Difference maps between the $\psi_E$ lensed and unlensed
  fields, obtained in the MS-modified-LP case (upper panel) and in the
  unmodified-LP case (lower panel). Units in $mK$.}
\label{psiE_diff_maps}
\end{center}
\end{figure*}
\begin{figure*}
\begin{center}
\resizebox{13cm}{!}{\includegraphics{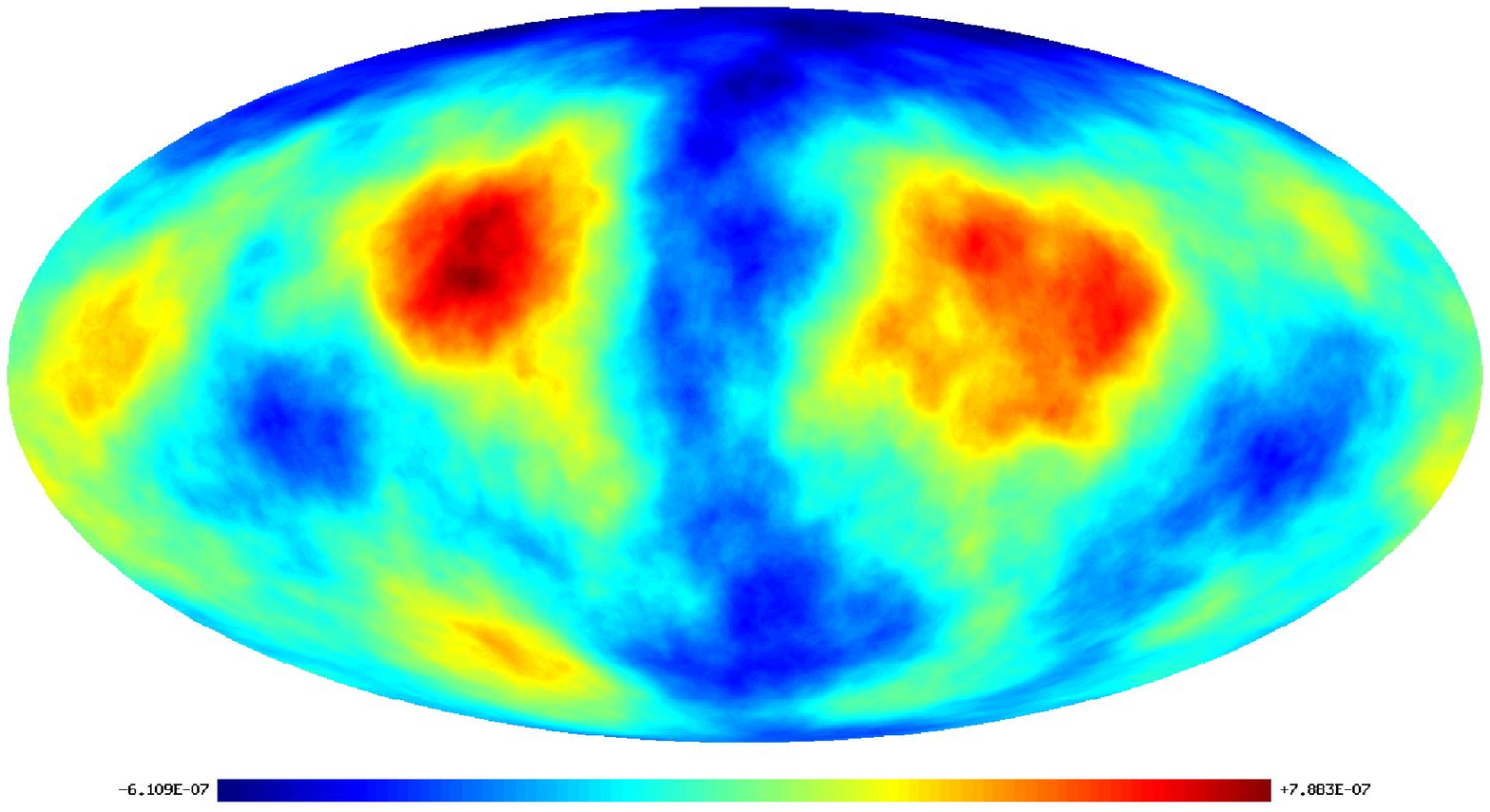}}
\resizebox{13cm}{!}{\includegraphics{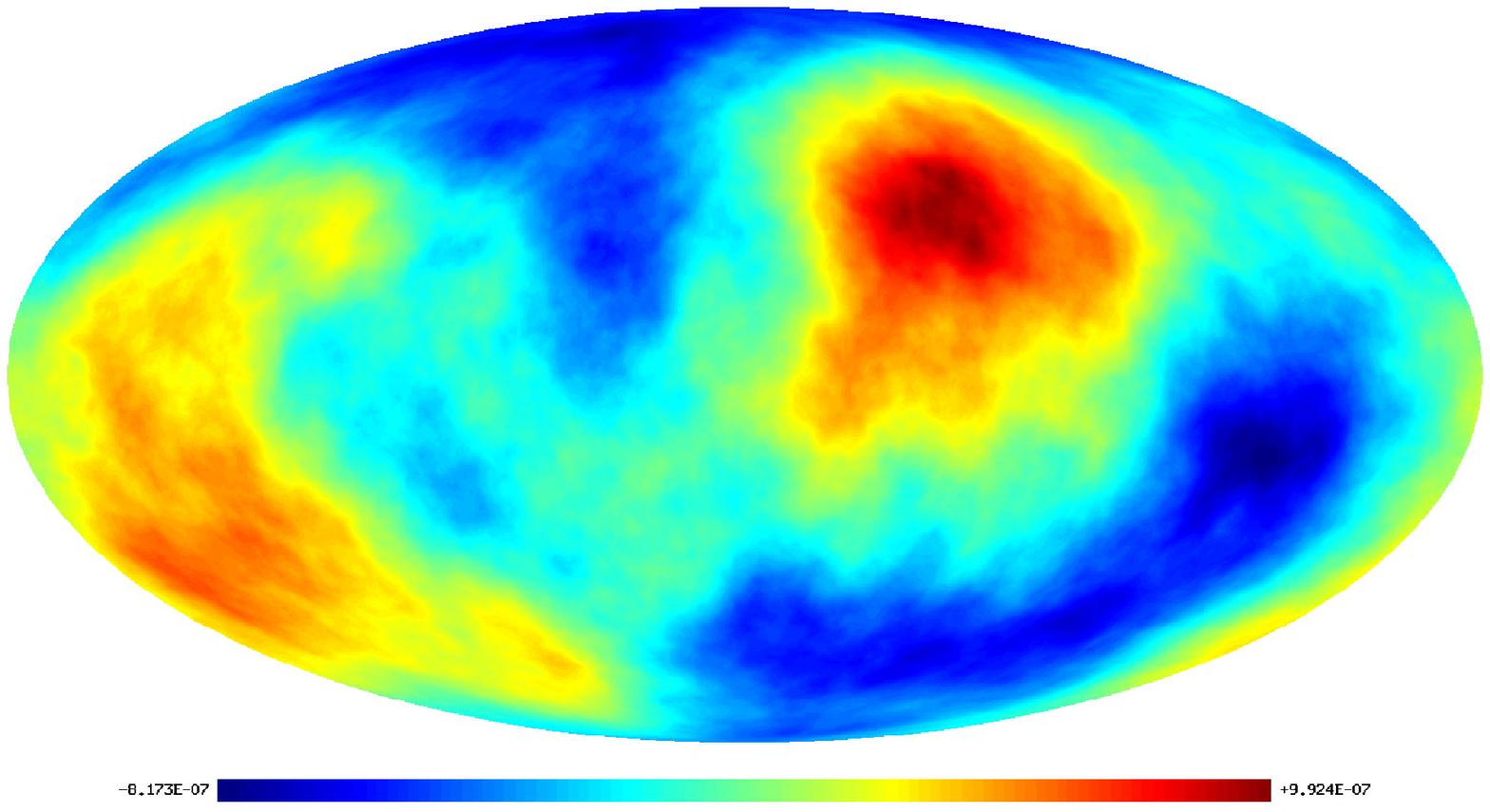}}
\footnotesize
\caption{Difference maps between the $\psi_B$ lensed and unlensed
  fields, obtained in the MS-modified-LP case (upper panel) and in the
  unmodified-LP case (lower panel). Units in $mK$.}
\label{psiB_diff_maps}
\end{center}
\end{figure*}
\begin{figure}
\includegraphics[width=.48\textwidth]{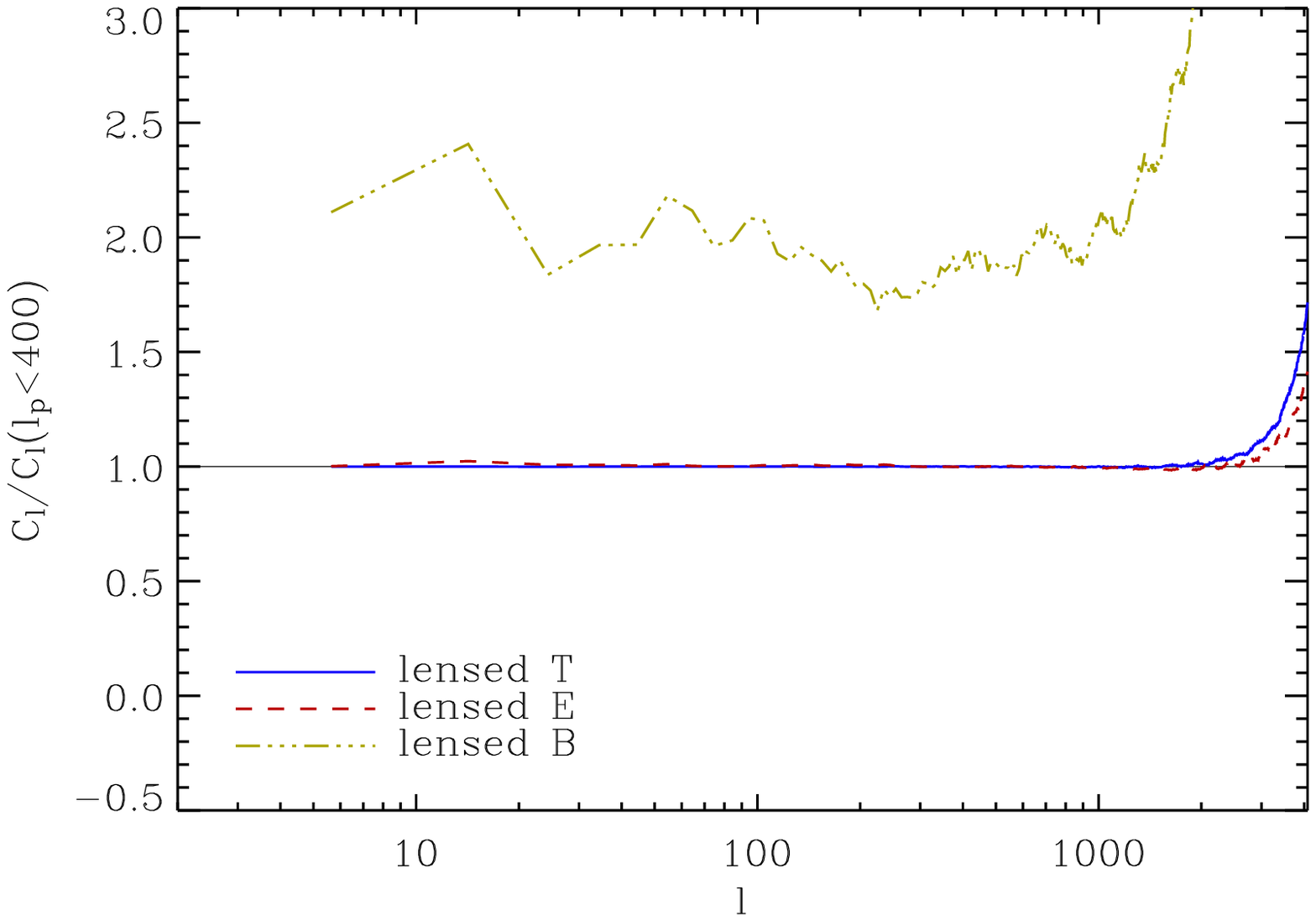}
\includegraphics[width=.48\textwidth]{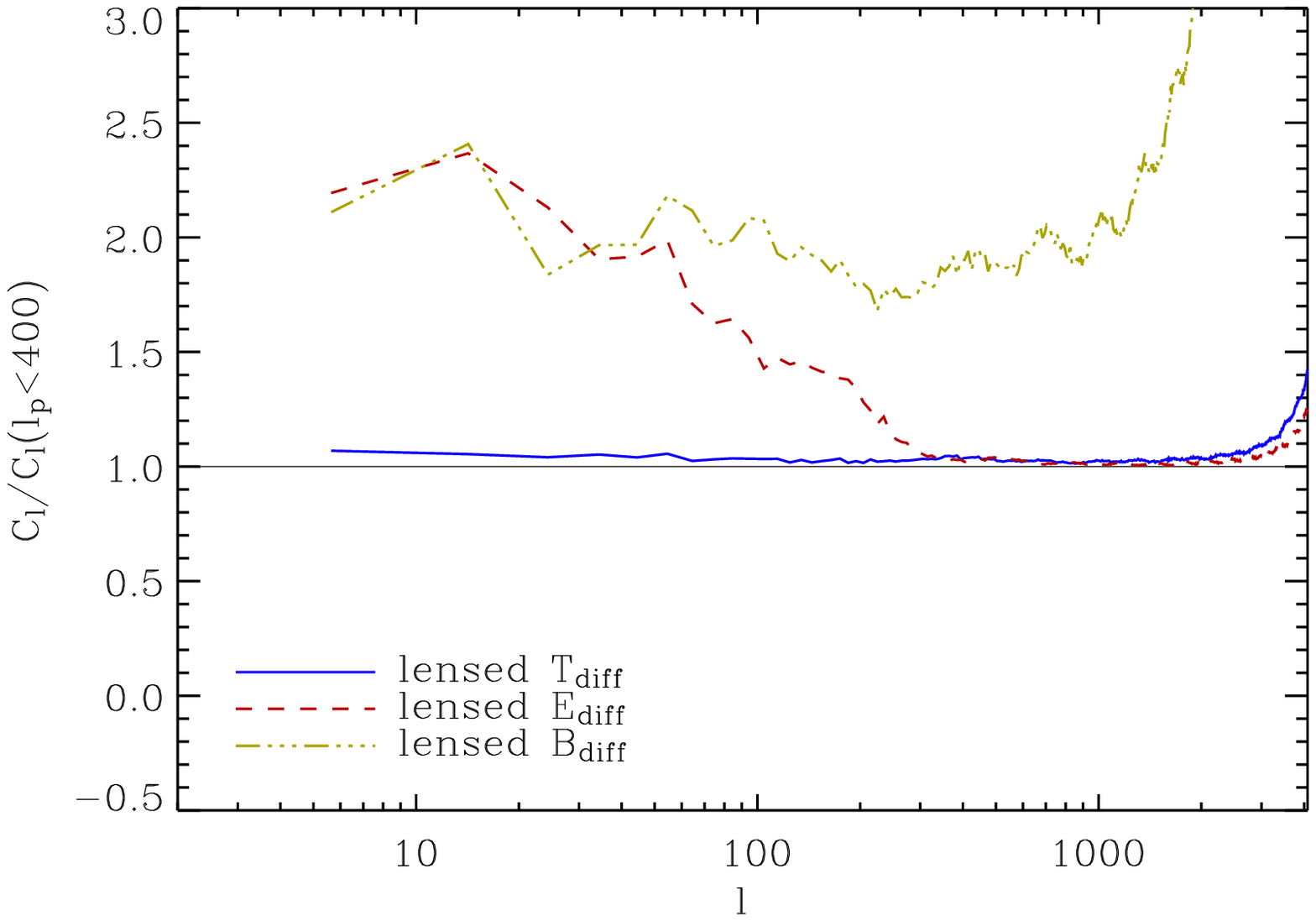}
\footnotesize
\caption{{\em Top panel:} The dot-dashed light green line represents
  the ratio between the lens-induced $B$ power spectrum, which is
  obtained using the multipole range $0\leq l_{lp}\leq 6143$ for the
  spherical harmonic coefficients of the lensing potential, and the
  spectrum obtained using only $0\leq l_{lp}\leq 400$. The solid blue
  line and the dashed red line represent these ratios in the $T$ and
  $E$ cases, respectively.  {\em Bottom panel:} The same as the upper
  panel when we calculate these ratios for the temperature and
  polarization power spectra extracted from the difference $T$,
  $Q$ and $U$ maps, in the $0\leq l_{lp}\leq 6143$ and
  $0\leq l_{lp}\leq 400$ cases, respectively.}
\label{Nol400}
\end{figure}
We emphasise that we always run CAMB using the same cosmological
parameters as the MS specified in Section 2, and, for consistency with
the MS map-making procedure, we have fixed the maximum redshift of the
line-of-sight integration for the CAMB lensing sources at $z_{\rm
  max}=127$. Indeed, the lensing power from even higher redshifts is
negligible for CMB lensing, as we show in Fig.~\ref{projpot_PS} where
the light-green long-dashed line overlaps the black dashed-dotted line
perfectly.

On multipoles $l > 400$, the cross-correlation between the
lensing-potential and the temperature is negligible, even if there
could be some residual contribution coming from the non-linear
Rees-Sciama effect. In this work we do not consider the
cross-correlation due to this second order effect.

To generate the lensed $T$, $Q$, $U$ fields from the MS-modified-LP
code, we adopt the interpolation
scheme described in Appendix E~4 of \cite{Lewis_E4}, using a high value of the multipole
$l_{\rm max}$ to maximize the accuracy.  This allows running the
simulation several times without excessive consumption of CPU time and
memory.  We work under the null hypothesis that tensor modes are
absent in the early Universe, so that the produced B-mode polarization
is due only to the power transfer from the primary scalar E-modes into
the lens-induced B-modes.  We choose $l_{\rm max}=6143$ and interp-factor $= 1.5$, 
effectively the same resolution as
HEALPix\footnote{http://healpix.jpl.nasa.gov/} with pixelization parameter
$N_{\rm side}=2048$, which corresponds to an angular resolution of
$\sim 1.72^\prime$ \citep{Healpix}, with $12 N_{\rm side}^2$ pixels in
total.

For comparison and testing, we also generate the corresponding
unmodified-LP and unlensed $T$, $Q$, $U$ maps, by choosing the same
positive seed in the LP parameter file (hereafter ``unmodified-LP''
stands for the results obtained by lensing the sky with the original
unmodified version of the LensPix code).

\section{Simulated lensed maps}
The subtraction of the unlensed CMB maps from the corresponding lensed
CMB maps allows highlighting the dark-matter distribution which causes
the gravitational deflection of the CMB photons. In the upper and
middle panels of Fig.~\ref{TP_diff_maps} we show the resulting
difference temperature and polarization maps for a particular seed
choice and for the MS-modified-LP case only, since the arcminute-scale
differences with respect to the unmodified-LP case are not visible by
the naked eye.  It is worth noting how the distribution of the
deflection-angle modulus (lower panel) reflects itself in the
distribution of the temperature difference map (upper panel) and in
the distribution of the map of the polarization modulus $\Delta
P=\sqrt{\Delta Q^2+\Delta U^2}$ (middle panel), which has been
obtained from the difference $Q$ and $U$ maps. 
Unfortunately, since the primary unlensed CMB is unknown, the difference maps cannot be 
directly observed. Anyway the simulated difference maps can help to physically and visually 
understand how large scale correlations are imprinted on the CMB due to the large scale modes in
the deflection field, and to catch effects that are not observable due to primary unlensed CMB.

The temperature and polarization difference maps have the peculiarity
of including all the information inferred from weak-lensing on the
primary unlensed CMB.  Nonetheless, it is well known that the lensed
$B$-modes of polarization are more sensitive to the non-linear
evolution of the cosmic structures than the $T$, $Q$ and $U$ modes
separately \citep{Lewis05}.  The difference between the unmodified-LP
and MS-modified-LP cases lies exactly at the non-linear level, since
the lensing-potential realizations differ at multipoles $l > 400$ in
the spherical harmonic space for the two cases.  Moreover, from
Fig.~\ref{projpot_PS}, we observe that the MS lensing potential shows
an excess of power on $l>2500$ with respect to the CAMB approximation,
and, as already pointed out, the non-linear dark matter evolution may
enhance the level of non-Gaussianity present in the lensed CMB maps.

Consequently, in order to verify if the different distributions of the
lensing potential in the MS-modified-LP and unmodified-LP cases could
visibly affect the lensed polarization distributions, we have
constructed the full-sky maps of the scalar $\psi_E$ and pseudo-scalar
$\psi_B$ potentials, which are related to the $Q$ and $U$ Stokes
parameters as follows \citep{E_B_decomposition}:
\begin{eqnarray}
Q+iU= \ra\ra(\psi_E+i\psi_B)\;,
\label{psiE}
\end{eqnarray}
\begin{eqnarray}
Q-iU= \lo\lo(\psi_E-i\psi_B)\;, 
\label{psiB}
\end{eqnarray}
where the spin-raising $\ra$ and spin-lowering $\lo$ operators on the
sphere are defined as \citep{Penrose}
\begin{eqnarray}
\ra = -\sin^s\theta\,\left[{\partial\over\partial\theta}+
i\csc\theta\,{\partial\over\partial\phi}\right]\sin^{-s}\theta \;,
\end{eqnarray}
\begin{eqnarray}
\lo=-\sin^{-s}\theta\left[{\partial\over\partial\theta}-i\csc\theta
{\partial\over\partial\phi}\right]\sin^s\theta \;,
\end{eqnarray}
and $s$ is the spin of the function to which the operator is applied.
The quantities $\psi_E$ and $\psi_B$ are directly related to the
electric and magnetic types of polarization, since their spherical
harmonic coefficients can be written in terms of the harmonic
coefficients of the $E$- and $B$-modes, respectively
\begin{eqnarray}
\psi_E=-\sum_\lm [(l-2)!/(l+2)!]^{1/2} a_\elm \Yo_\lm \;,
\label{psiE_lm}
\end{eqnarray}
\begin{eqnarray}
\psi_B=-\sum_\lm [(l-2)!/(l+2)!]^{1/2} a_\blm \Yo_\lm \;.
\label{psiB_lm}
\end{eqnarray}

The $\psi_E$ and $\psi_B$ potentials are very useful for real space
calculations and, exploiting the HEALPix routine SYNFAST
\citep{Healpix}, we have produced the corresponding maps as synthetic
realizations, using the $E$, $B$ spherical harmonic coefficients
extracted (via the HEALPix routine ANAFAST) from the lensed $Q$, $U$
simulated maps and multiplied by the prefactors of
Eqs.~(\ref{psiE_lm})-(\ref{psiB_lm}), respectively.  On high
multipoles these prefactors have the asymptotic form
$[(l-2)!/(l+2)!]^{1/2}\sim l^{-2}$, consequently their effect is to
suppress the small scale power with respect to the $E$- and $B$-mode
cases, and to make the large scale structure differences much more
evident.

In fact, looking at Figs.~\ref{psiE_diff_maps}-\ref{psiB_diff_maps},
we observe that the lensed $\psi_E$ and $\psi_B$ difference maps have
a \emph{degree}-scale distribution which differs in the MS-modified-LP
and unmodified-LP cases, even if the corresponding lensing-potential
maps differ on \emph{arcminute} scales ($l>400$).  This result is more
expected in the $B$-mode case since it is well-known that the
non-linear density field introduces $\sim 10$\% corrections to the
lens-induced $B$-mode power on all scales \citep{Lewis06}, so that
multipoles $l>400$ will affect the $\psi_B$-field realizations on
larger scales too. The same is instead less obvious in the $E$-mode
case.

In order to understand the reason of the different large-scale
distribution of these maps, we have also produced the lensed
temperature, $\psi_E$ and $\psi_B$ fields using a Gaussian
lensing-potential distribution which includes only the scales that
correspond to multipoles up to $l_{lp}=400$ in the spherical harmonic
domain (where the subscript ``lp'' stands for lensing-potential). For
comparison among the figures, we have used the same seed as for the
fully ($0\leq l_{lp} \leq 6143$) lensed maps, even if the results have
been tested for different seeds.  In Fig.~\ref{Nol400} we show the
ratios between the signals corresponding to the two cases, both for
the lensed- and difference-maps.

In particular, if we consider the lensed $T$- and $E$-mode power
spectra (top panel of Fig.~\ref{Nol400}) which include the contribution from the primary
CMB, we do not observe any substantial difference on the large scales
between the two cases $0\leq l_{lp}\leq 400$ and $0\leq l_{lp}\leq
6143$, so that on large scales the lensed electric polarization
appears not to be much affected by the non-linear scales in the same
way as the lensed temperature is. The $B$ case is different since we
are working with the null hypothesis of vanishing primary magnetic
polarization, so that we are considering here only lens-induced
$B$-modes.

Nonetheless, if we consider the power spectra extracted from the
difference $T$ and $\psi_E$ maps (where we are subtracting the primary
CMB), and again take the ratios between the signals in the two cases
$0\leq l_{lp}\leq 400$ and $0\leq l_{lp}\leq 6143$ (lower panel of
Fig.~\ref{Nol400}), we see that on large scales the $E$-mode
polarization, cleaned of the primary signal, gets power from the
smaller scales. In particular we see that, in the $E$ and $B$ cases,
multipoles larger than $l_{lp}=400$ transfer about $100$\% of the
power to the low multipoles $l\lesssim 100$, i.e. on
scales large enough for the flat-sky approximation to be inadequate.  
This transfer of power towards low multipoles is possible only for peculiar alignment angles
between the unlensed $E$-mode and the lensing structures, i.e.  for
almost, but not quite aligned modes. This is analogous to the
long-wavelength beat mode one obtains from superposing two
oscillations with two almost equal frequencies.  We do not observe the
same effect when we consider the ratio between the signals extracted
from the temperature difference maps in the two cases $0\leq
l_{lp}\leq 400$ and $0\leq l_{lp}\leq 6143$, where the transfer of
power is less than $10$\% on $l\lesssim 100$.  As we move to larger
multipoles instead, the power transfer decreases and, for $ 400 \leq l
\leq 2000$, the lensed $E$ trend starts converging to the lensed
temperature one, while the power of the lens-induced $B$-modes goes on
increasing with the multipoles, still fed by the non-linear scales in
the lensing-potential at $l_{lp}>400$.

Therefore we conclude that in
Figs.~\ref{psiE_diff_maps}-\ref{psiB_diff_maps} the different
degree-scale distributions which characterize the MS-modified-LP and
unmodified-LP $\psi_E$ ($\psi_B$) difference map realizations (even
though the corresponding lensing-potential realizations differ only on
$l_{lp}>400$) represent the imprints of the power transfer from
smaller to larger scales induced by the lensing remapping onto the
lensed $E$ and $B$ signals cleaned of the primary CMB, in a way which
differs from what happens to the lensed temperature.
\begin{figure}
\begin{minipage}{1.0\linewidth}
\centering
\vspace{0.5cm}
\includegraphics[width=0.9\textwidth]{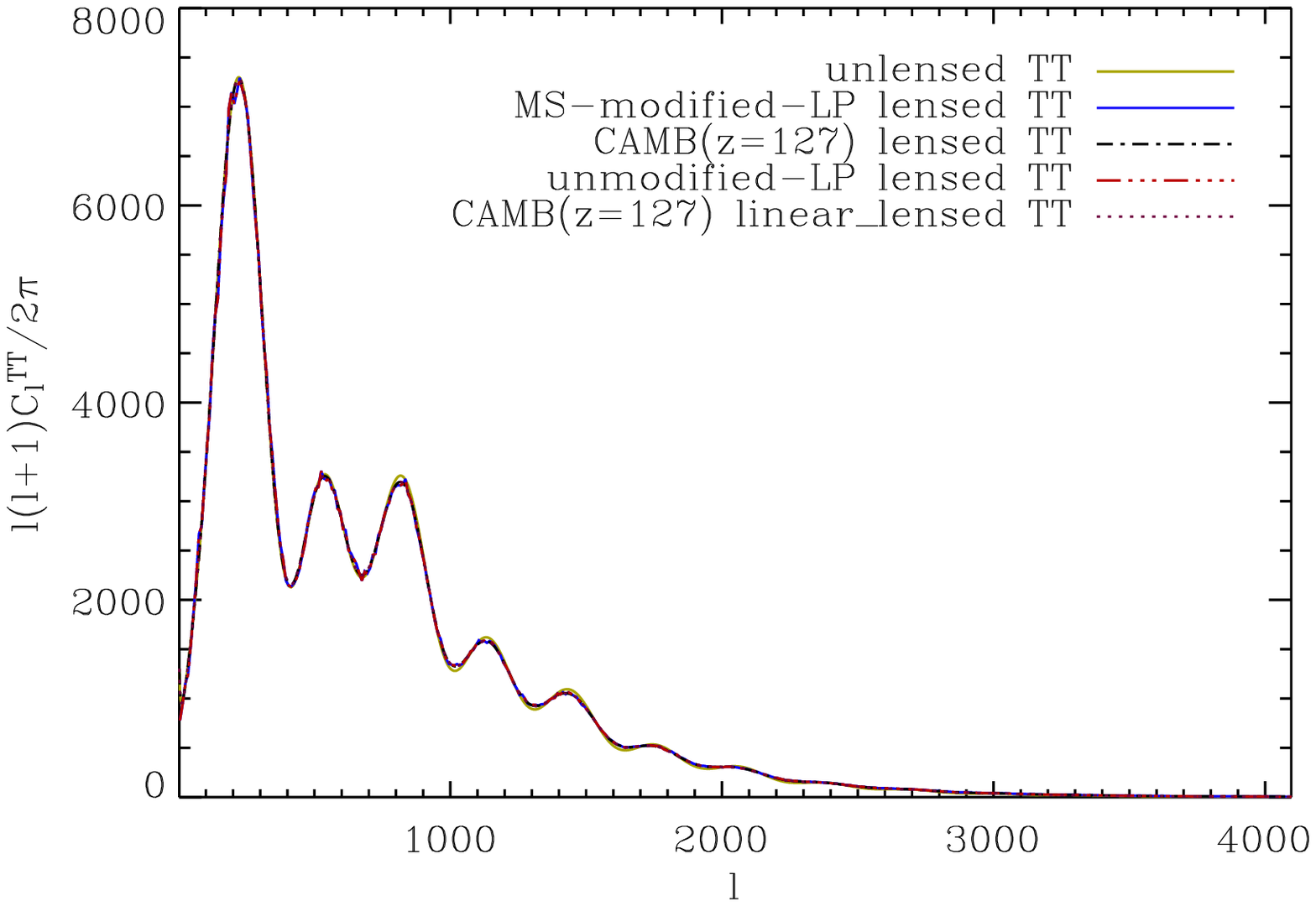}
\vspace{1.5cm}
\end{minipage}
\begin{minipage}{1.0\linewidth}
\centering
\includegraphics[width=0.83\textwidth]{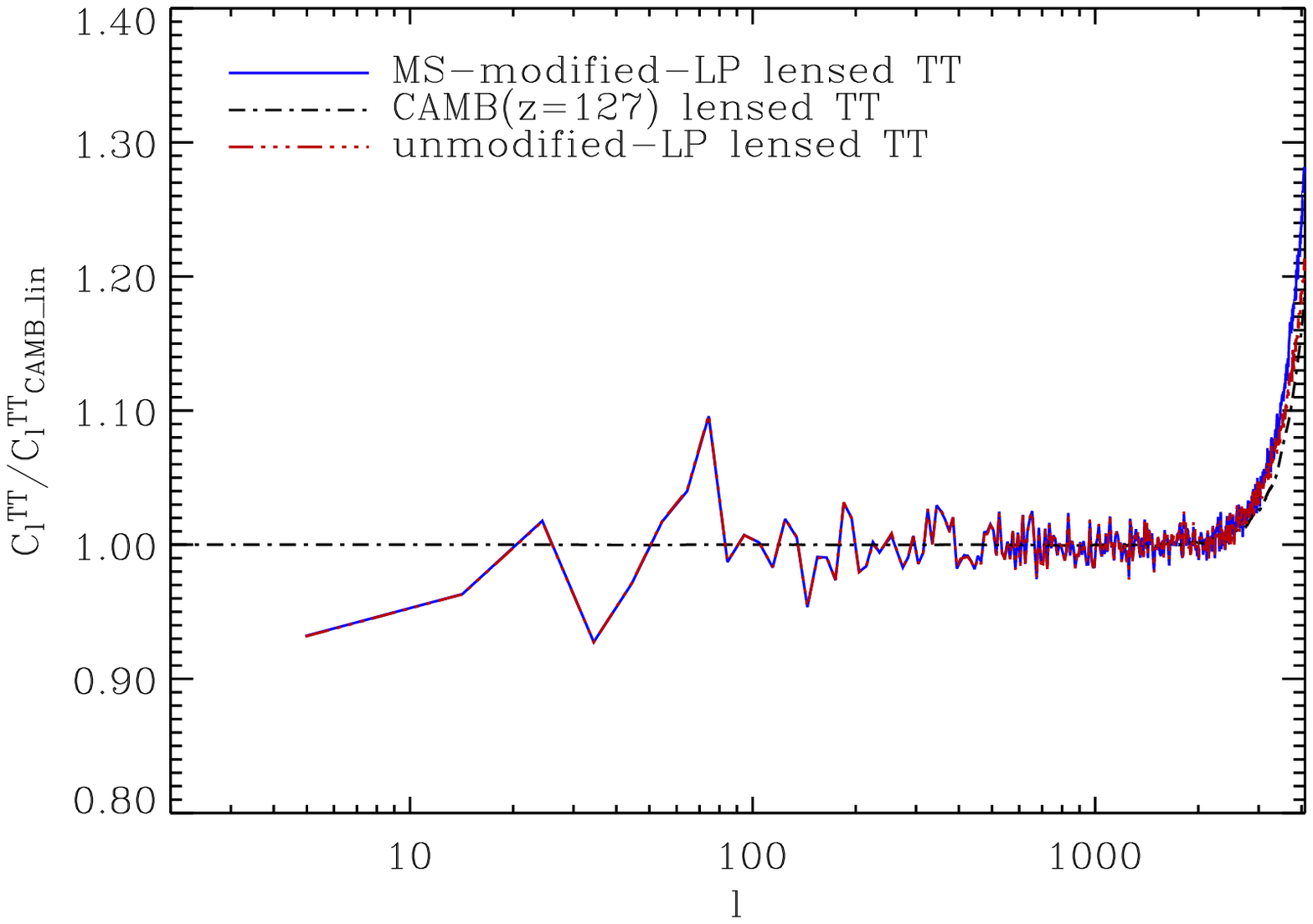}
\end{minipage}
\vspace{1.5cm}
\footnotesize
\caption{{\em Top panel:} Temperature power spectra ($\mu K^2$) for
  the different cases described in the text.  {\em Bottom panel:}
  Temperature power spectrum ratios with respect to the linear lensed
  case.}
\label{TT_PS}
\end{figure}
\begin{figure}
\begin{minipage}{1.0\linewidth}
\centering
\vspace{0.5cm}
\includegraphics[width=0.9\textwidth]{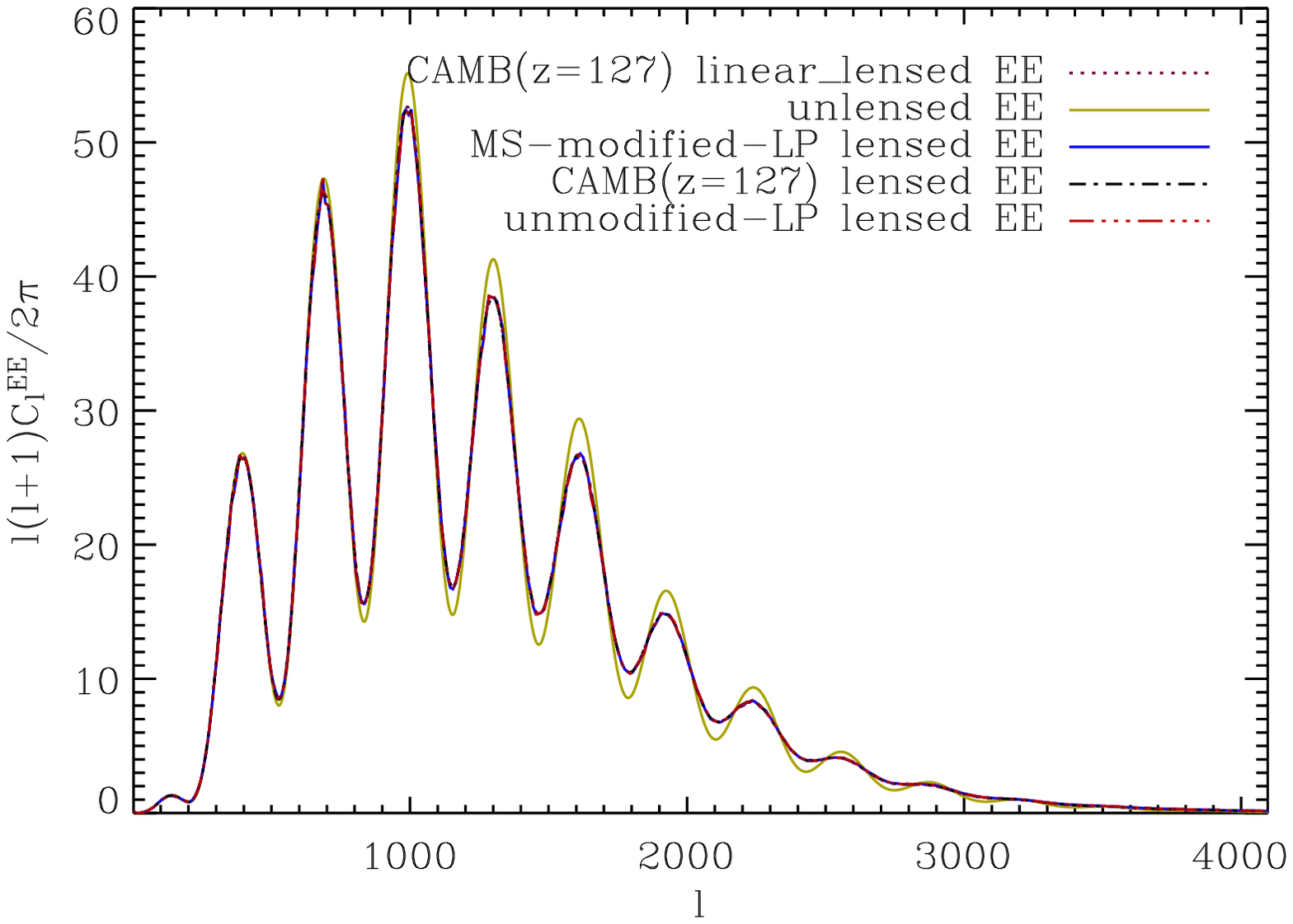}
\vspace{1.3cm}
\end{minipage}
\begin{minipage}{1.0\linewidth}
\centering
\includegraphics[width=0.83\textwidth]{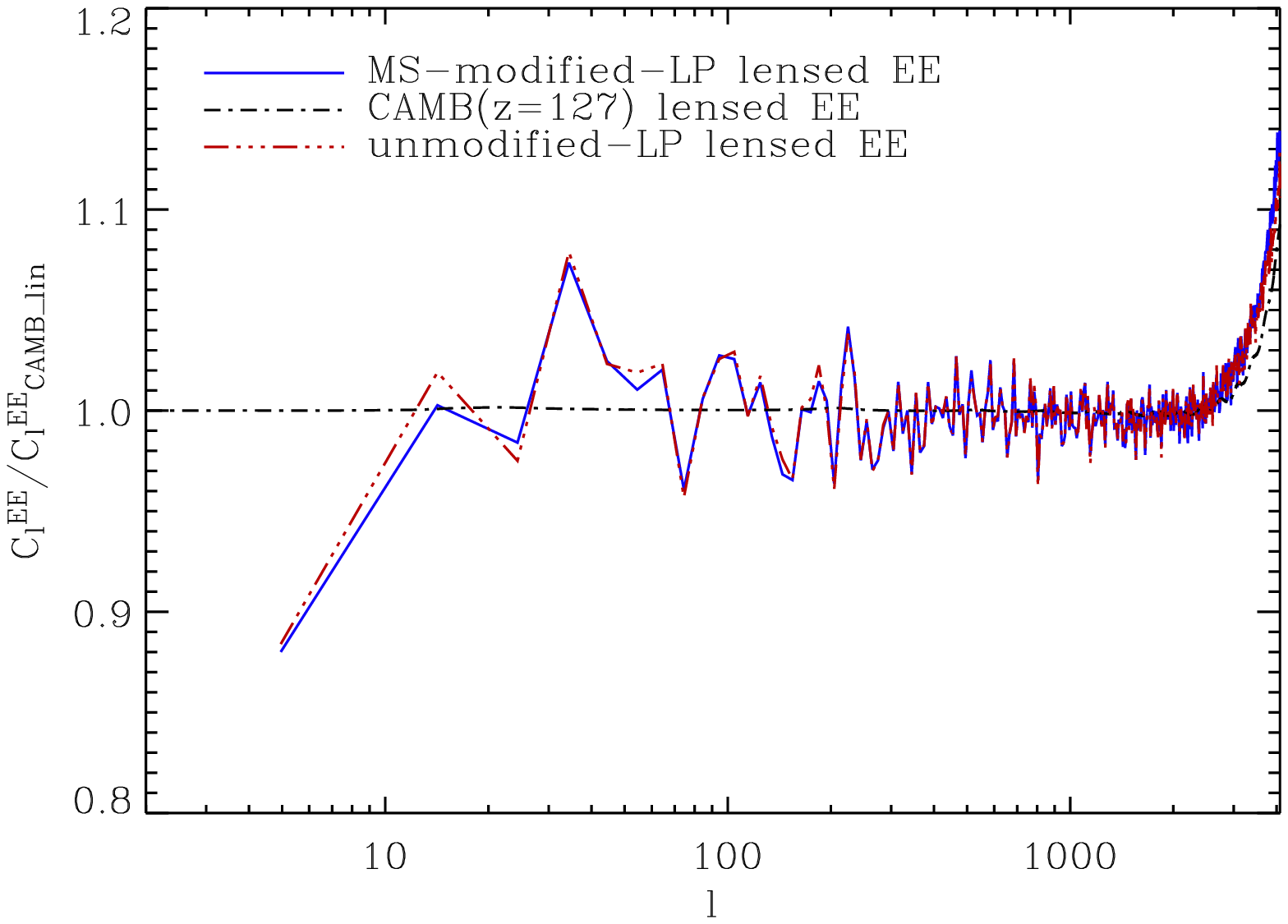}
\end{minipage}
\vspace{1.5cm}
\footnotesize
\caption{{\em Top panel:} $E$-mode power spectra ($\mu K^2$) for the
  different cases described in the text.  {\em Bottom panel:} $E$-mode
  power spectrum ratios with respect to the linear lensed case.}
\label{EE_PS}
\end{figure}
\begin{figure}
\begin{minipage}{1.0\linewidth}
\centering
\vspace{0.5cm}
\includegraphics[width=0.9\textwidth]{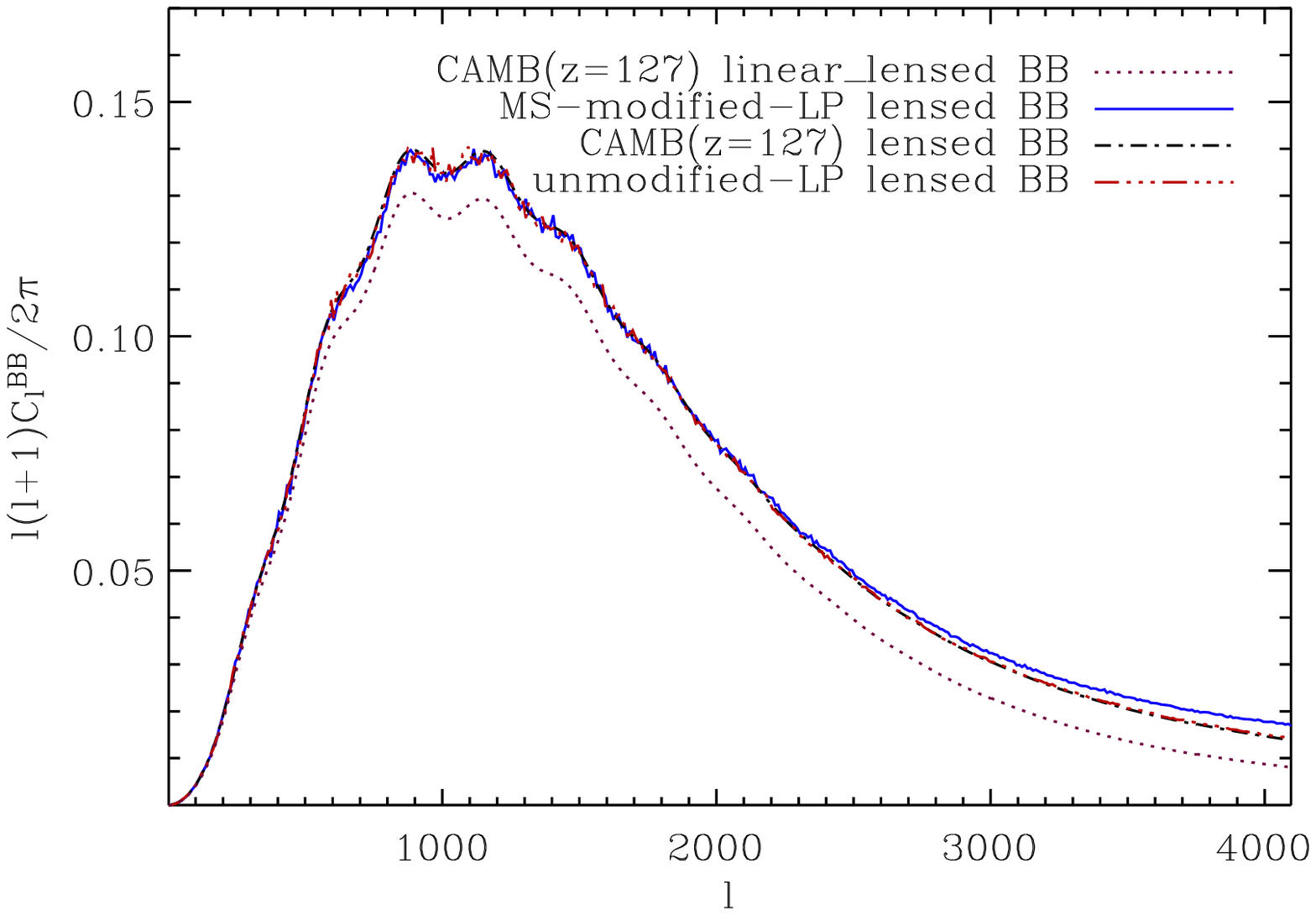}
\vspace{1.0cm}
\end{minipage}
\begin{minipage}{1.0\linewidth}
\centering
\includegraphics[width=0.83\textwidth]{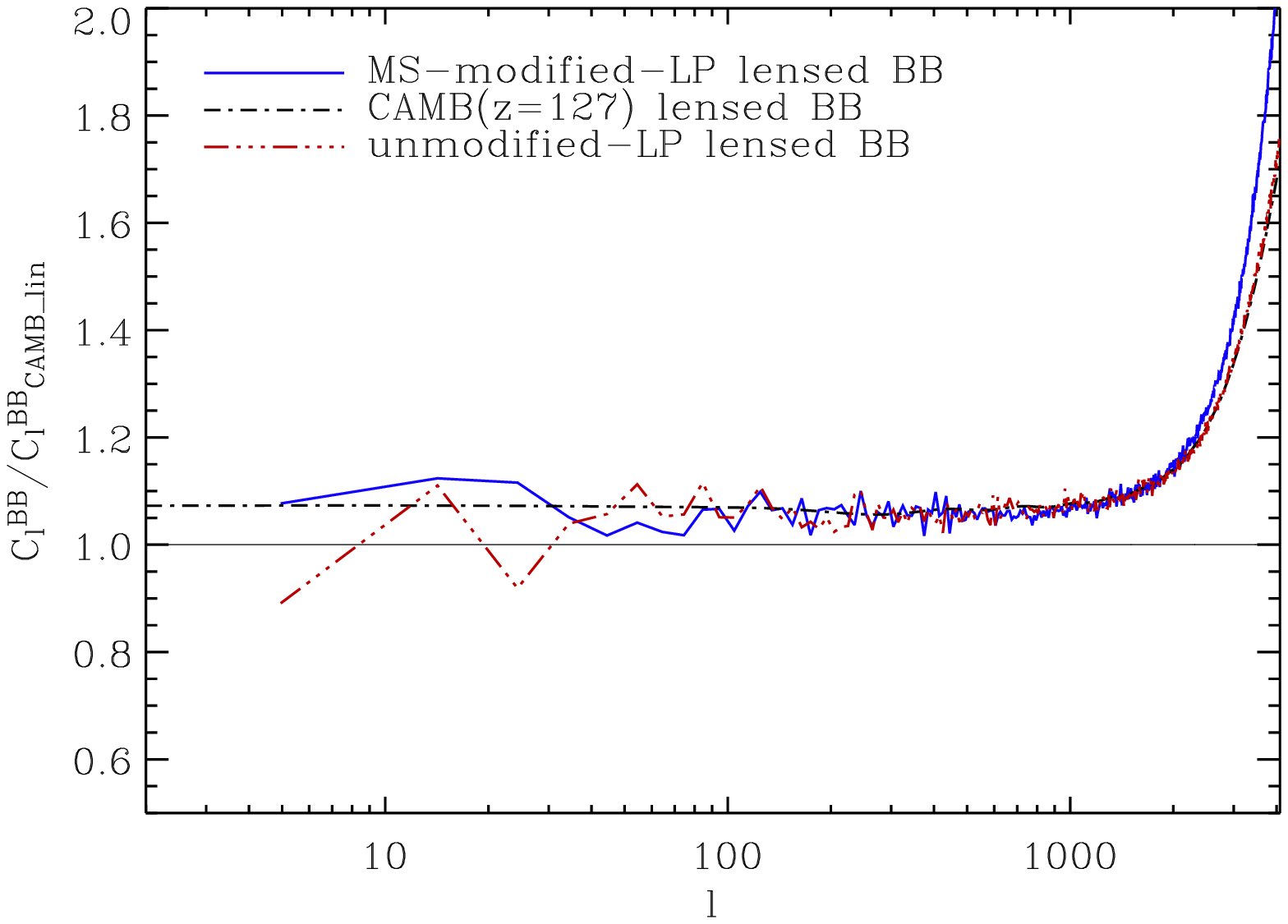}
\end{minipage}
\vspace{2cm}
\footnotesize
\caption{{\em Top panel:} Lens-induced $B$-mode power spectra ($\mu
  K^2$) for the different cases described in the text.  {\em Bottom
    panel:} Lens-induced $B$-mode power spectrum ratios with respect
  to the linear lensed case.}
\label{BB_PS}
\end{figure}
\begin{figure}
\begin{minipage}{1.0\linewidth}
\centering
\vspace{0.5cm}
\includegraphics[width=0.9\textwidth]{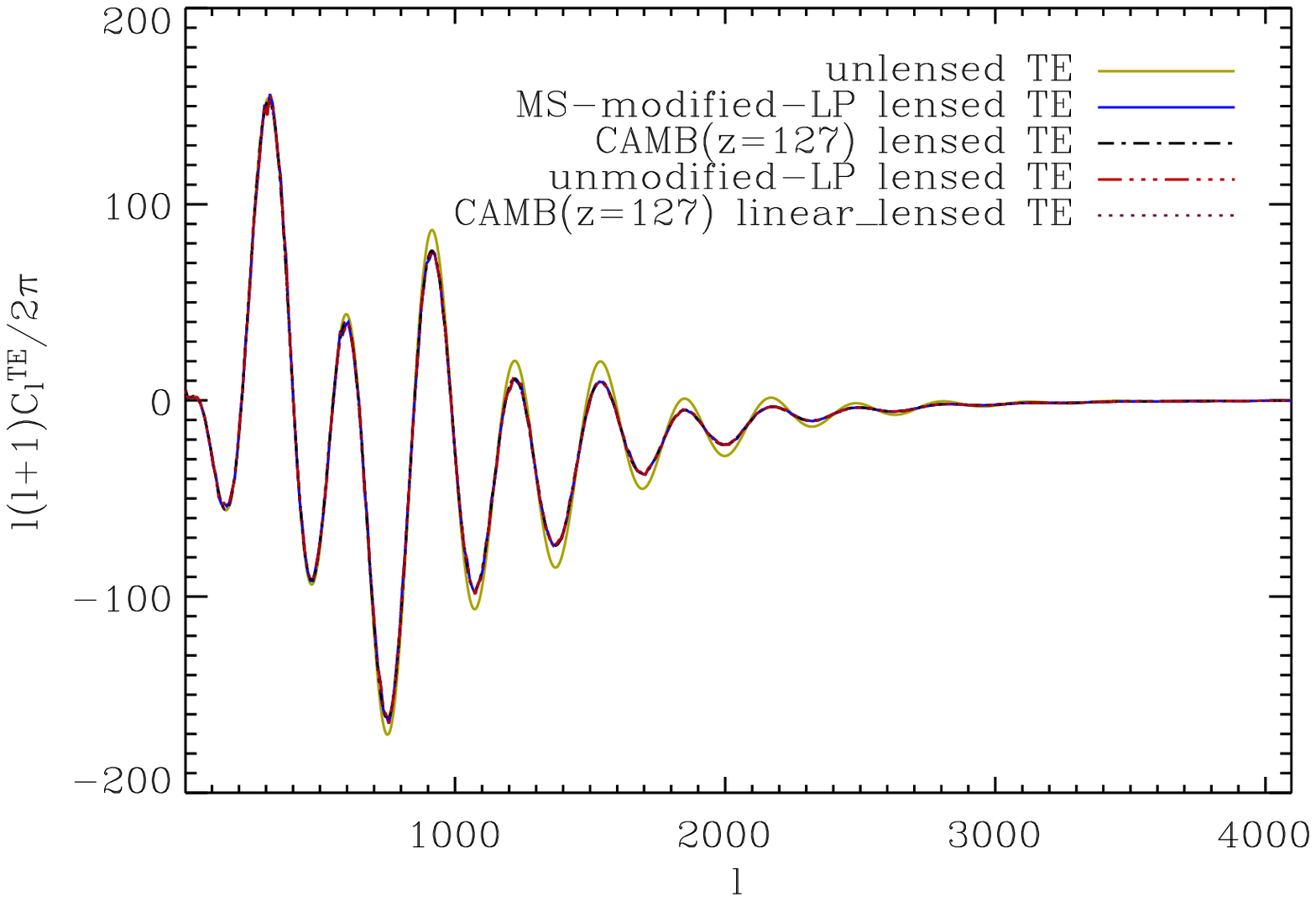}
\vspace{3.2cm}
\end{minipage}
\begin{minipage}{1.0\linewidth}
\centering
\includegraphics[width=0.83\textwidth]{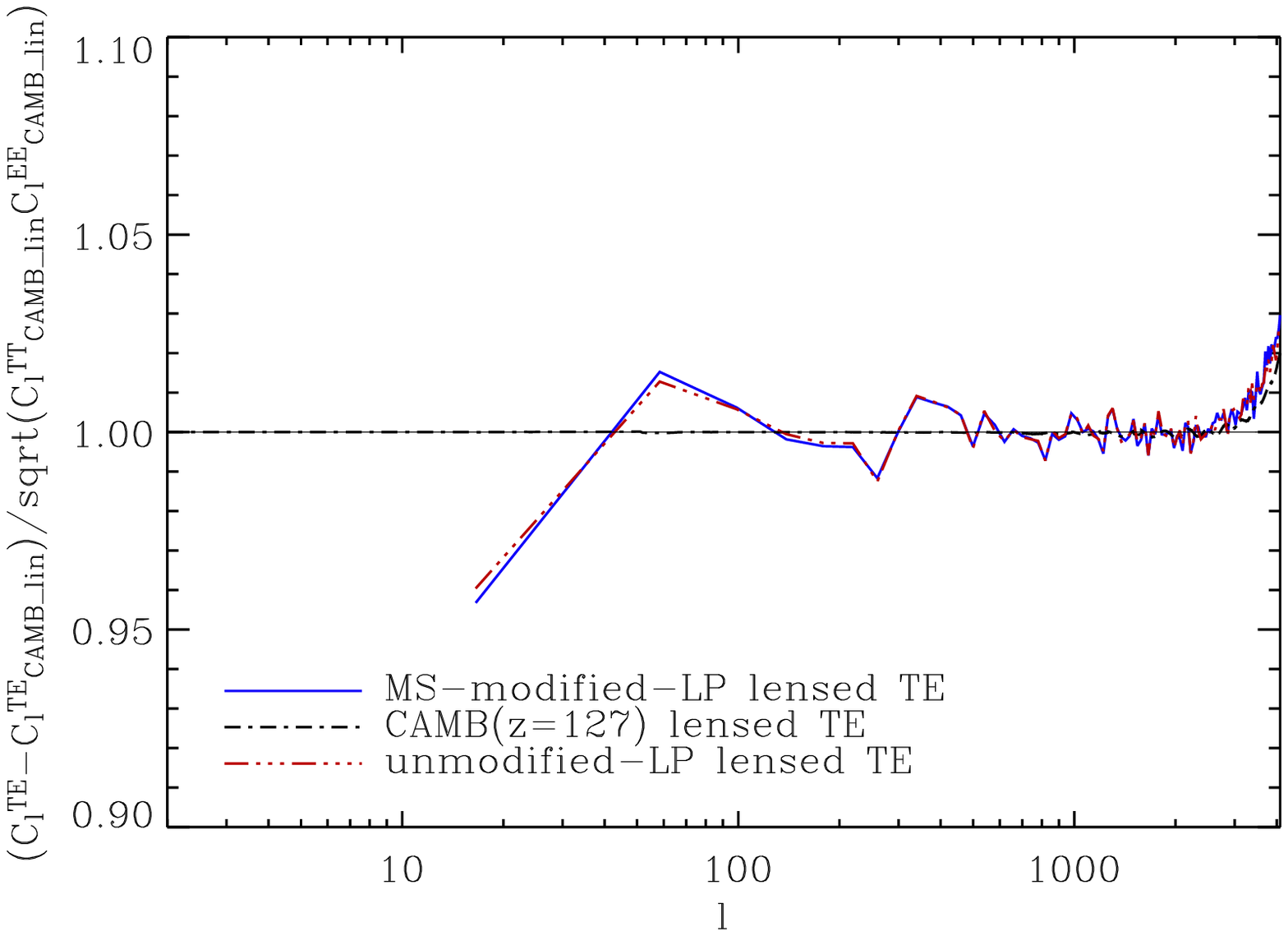}
\end{minipage}
\vspace{1.2cm}
\footnotesize
\caption{{\em Top panel:} $T$-$E$ cross-power spectra ($\mu K^2$) for
  the different cases described in the text.  {\em Bottom panel:}
  The fractional change in the lensed $T$-$E$ cross-power spectrum due to the non-linear matter 
  evolution for the different cases described in the text.}
\label{TE_PS}
\end{figure}
\begin{figure}
\includegraphics[width=.48\textwidth]{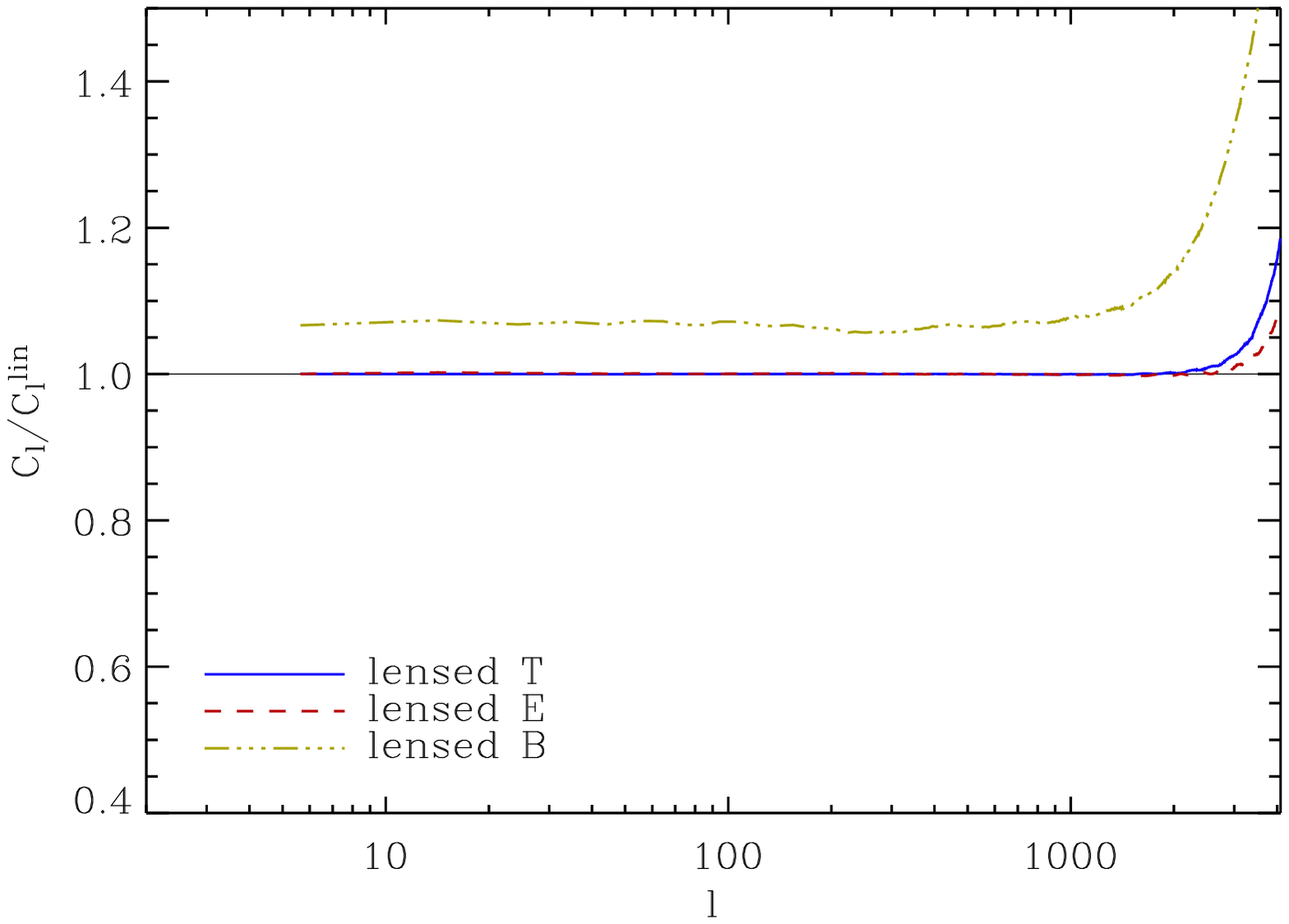}
\includegraphics[width=.48\textwidth]{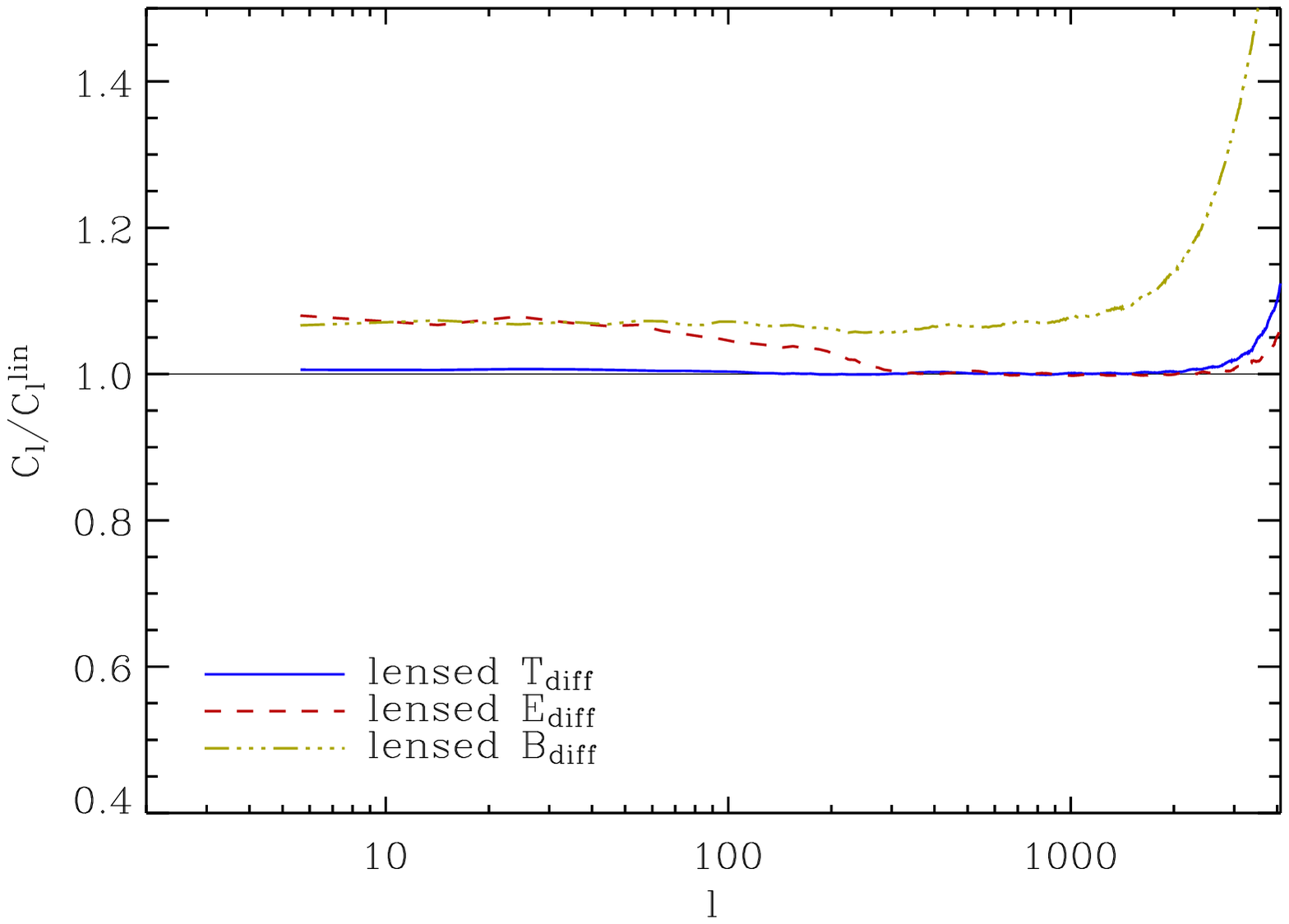}
\footnotesize
\caption{{\em Top panel:} The dot-dashed light green line represents
  the ratio between the lens-induced $B$ power spectrum, which is
  obtained including the non-linear cosmic structure evolution, and
  the spectrum obtained in the linear limit.  The solid blue line and
  the dashed red line represent these ratios in the $T$ and $E$ cases,
  respectively.  {\em Bottom panel:} The same as the upper panel when
  we calculate these ratios for the temperature and polarization power
  spectra extracted from the difference $T$, $Q$ and $U$
  maps, in the linear and non-linear cases, respectively.}
\label{linearNOlinear}
\end{figure}

\section{Angular power spectra}
In this Section we perform several quantitative analyses of the
obtained results.  Our principal aim is to test the consistency and
quantify the differences between the theoretical expectations and the
findings from our simulation procedure.

As a first check, we have extracted from the corresponding maps the
angular power spectra of the MS-modified-LP and unmodified-LP
simulated lensed $T$-, $E$- and $B$-components, together with the $TE$
lensed cross-correlation power spectrum. This is simply done with the
use of ANAFAST, adopting the correct deconvolution rules for the
HEALPix pixel window functions, and with
Eqs.~(\ref{psiE_lm})-(\ref{psiB_lm}) relating the spherical harmonic
coefficients of the $\psi_E$ and $\psi_B$ potentials to the
corresponding coefficients of the $E$- and $B$-modes.

We have also checked that the MS-modified-LP and unmodified-LP
simulated spectra extracted directly from the maps agree with the ones
obtained as direct outputs of the MS-modified-LP and unmodified-LP
codes, respectively.  Moreover, to test the accuracy, we have compared
all the angular power spectra with the lensed CMB power spectra
obtained using the all-sky correlation function technique implemented
in CAMB \citep{Challinor&Lewis05}.  These angular power spectra are
shown in the upper panels of Figs.~\ref{TT_PS}-\ref{TE_PS}
respectively, together with the CAMB unlensed and the linearly lensed
spectra, where the latter are obtained in the linear approximation of
the cosmic structure growth. Moreover, in the lower panels of the same
figures we show the ratios between the simulated lensed CMB angular
spectra with respect to the corresponding linear signal.

In the $TT$, $EE$ and $TE$ cases, it is clearly visible that CMB
lensing smears out the acoustic peaks by transferring power from
larger to smaller scales.  Moreover, since we are working under the
null hypothesis of vanishing intrinsic tensor modes, Fig.~\ref{BB_PS}
represents the power spectrum of the lens-induced $B$-modes into which
part of the primary $E$-modes has been converted as a result of the
displacements and distortions induced from the gravitational
deflection onto the electric-type polarization field.

As Figs.~\ref{TT_PS}, \ref{EE_PS} and \ref{TE_PS} show, for all the
lensed $TT$, $EE$ and $TE$ angular spectra, and for all multipole
orders up to $l\le 3500$, we observe a mostly perfect agreement
between the MS-modified-LP simulated signals and the unmodified-LP and
CAMB ones, where the non-linear structure evolution is
semi-analytically taken into account.  On the very large multipoles
$l>3000$, a small difference appears between the CAMB and
unmodified-LP lensed $TT$, $EE$ and $TE$ spectra, probably due to a
numerical effect deriving from the different computational machinery
implemented in CAMB and LensPix respectively.  The non-linear effects
start to be important at $l>2500$, according to the semi-analytical
expectations, and, on multipoles $l\sim 4000$, which correspond to
angular scales of few arcminutes, they grow up to $\sim$20\% for the
lensed temperature and up to $\sim$10\% for the lensed electric
polarization. Moreover, the excess of power present in the MS
lensing-potential power spectrum at multipoles $l \ge 2500$ (see
Fig.~\ref{projpot_PS}) manifests itself as a slight excess of power in
the MS-modified-LP lensed $TT$ and $EE$ spectra at $l \ge 3500$.  In
particular, at $l=4096$, the non-linearities present in the Millennium
Simulation produce a $\sim$6\% excess in the lensed temperature and a
$\sim$1.3\% excess in the lensed electric polarization, with respect
to the unmodified-LP case.

On the other hand, the excess of power due to the MS non-linearities
is much more evident in the angular power spectrum of the magnetic
component (see Fig.~\ref{BB_PS}) where this effect starts already at
$l\sim 2500$ and grows up to $\sim$14\% at $l=4096$, with respect to
the unmodified-LP case.  Moreover, the non-linear effects, which in
the lensing potential appear at $l \ge 400$, spread on all the scales
of the lens-induced $B$-mode spectrum, being already of the order of
$\sim$7.5\% on all the multipoles $l \leq 1000$, and growing up to
more than $\sim$70\% at $l=4096$.  As we have previously noticed, this
non-linear effect on the magnetic polarization is simply explained
considering that the lens-induced $B$-modes are very sensitive to the
non-linear evolution of the cosmic structures.

For what concerns the non-linear effects, as noticed already in
Sec.~3, it is important to stress again that, when considering the
power spectra extracted from the difference maps (in which the primary
CMB has been subtracted), we notice a very similar large-scale
behavior between the electric and magnetic polarization, which
strongly differs from the temperature trend. In order to analyse this
effect, we have produced the \emph{linearly}-lensed $T$, $Q$, $U$ maps
and subtracted the unlensed field.  In Fig.~\ref{linearNOlinear} we
show the two cases: the ratios of the lensed $T$, $E$, $B$ spectra
with respect to the corresponding spectra extracted from the
linearly-lensed maps, and the same ratios extracted from the
corresponding lensed and linearly-lensed \emph{difference} maps. While
in the first case (top panel) the transfer of power from small scales
to large scales is only visible in the lens-induced $B$-mode spectrum,
in the second case (bottom panel) this power transfer is also
observable in the lensed $E$-mode spectrum, with the same amount of
$\sim$7.5\% up to $l\lesssim 100$ as for the lens-induced $B$-modes.
The same does not occur to the lensed temperature even after
subtracting the unlensed CMB.  Finally, moving to higher multipoles,
the $E$ and $T$ signals converge to the same trend, while the $B$
signal keeps on growing because of the non-linear power.

\section{One-point statistics}

\begin{figure}
\begin{minipage}{1.0\linewidth}
\centering
\vspace{0.7cm}
\includegraphics[width=0.9\textwidth]{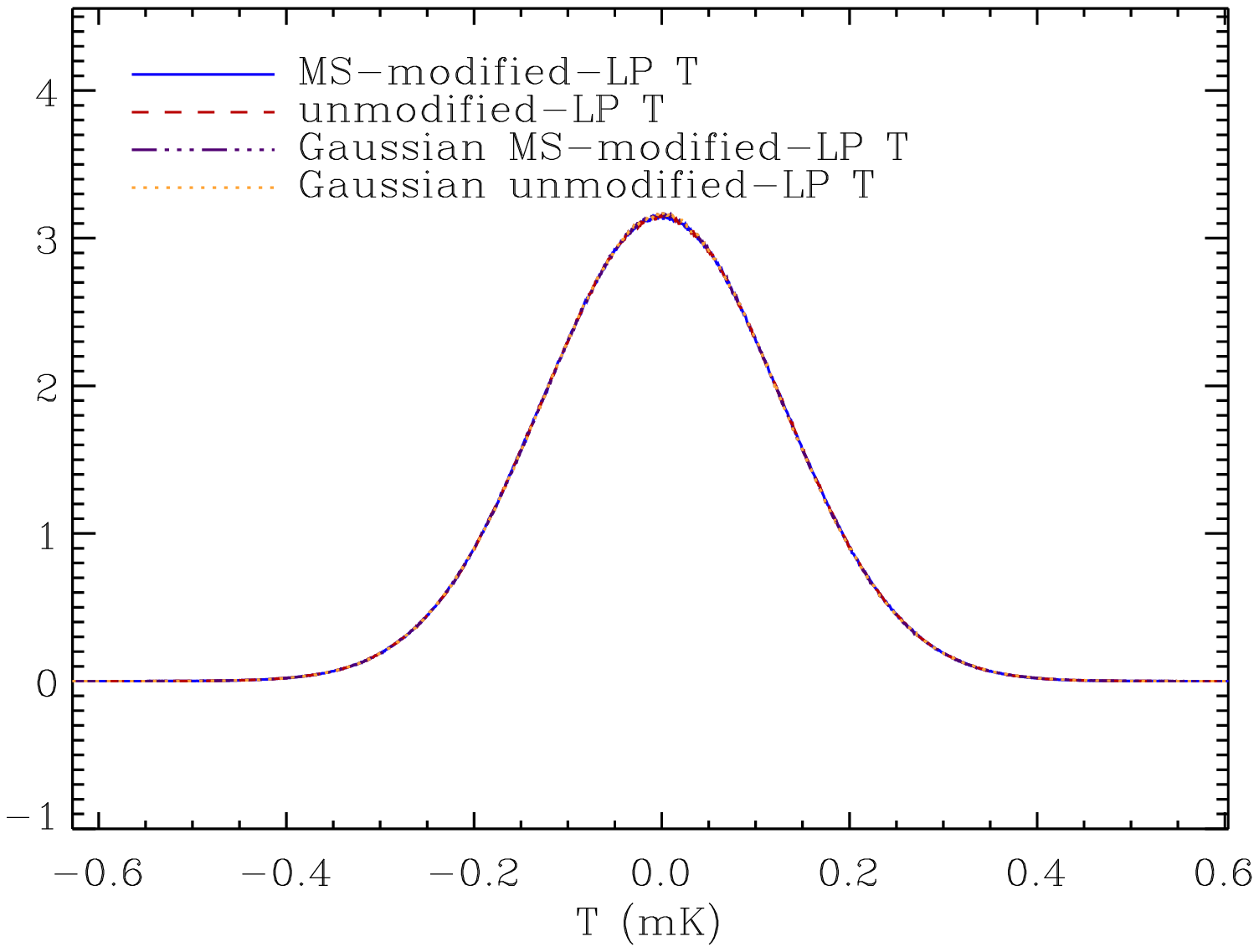}
\vspace{2.5cm}
\end{minipage}
\begin{minipage}{1.0\linewidth}
\centering
\includegraphics[width=0.9\textwidth]{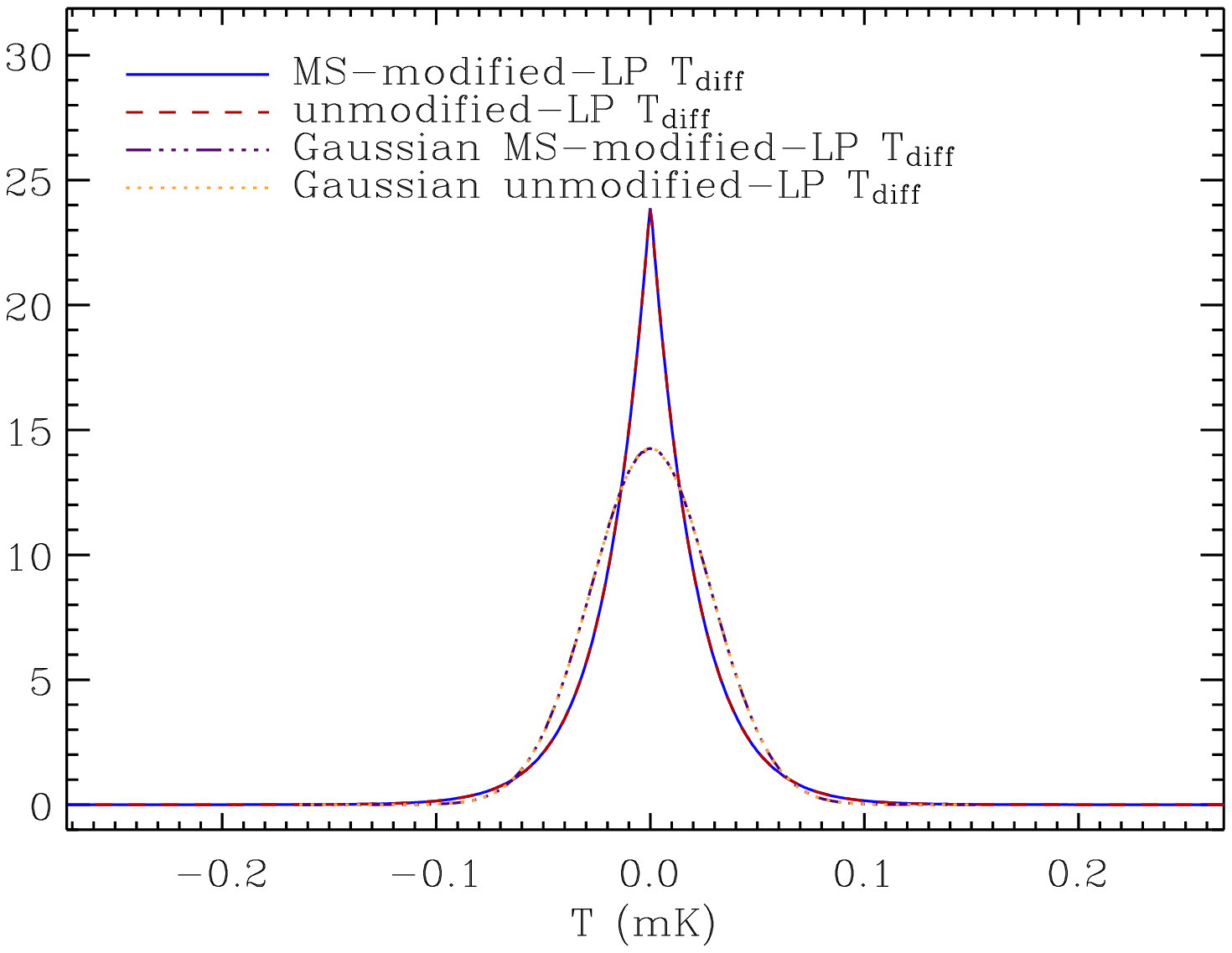}
\end{minipage}
\vspace{1cm}
\footnotesize
\caption{{\em Top panel:} The PDF of the temperature maps obtained in
  the MS-modified-LP and unmodified-LP cases compared to the PDF of
  Gaussian distributions with the corresponding mean value and
  standard deviation, respectively. {\em Bottom panel:} The same as in
  the upper panel for the PDFs of temperature difference maps.}
\label{TT_pdf}
\end{figure}

As a second analysis, we compare the one-point probability density
distributions (PDFs) of the lensed $T$, $Q$, $U$ maps with the PDFs of
Gaussian distributions randomly generated using the corresponding mean
and standard deviation values of the lensed maps.  The PDFs of both
the simulated MS-modified-LP and unmodified-LP $T$, $Q$, $U$ maps do
not show evidence of non-Gaussian features (upper panel of
Fig.~\ref{TT_pdf}), and are consistent with the unlensed temperature
and polarization distributions within a few per cent. This occurs thanks to the high
angular resolution of our maps (see e.g. \citet{cumulants}),
and is understandable because the main effect of CMB lensing is to transfer
power among different scales, without generating new power (and we
assume that the primary unlensed CMB field has a Gaussian
distribution). 

On the other hand, it is well-known that the remapping induced by
lensing onto the $T$, $Q$, $U$ fields generates a non-Gaussian
signature in the lensed sky which is optimally characterized by higher
order statistics as the bispectrum and trispectrum \citep[see
  e.g.][and references therein]{Lewis06}. This is mainly due to the
fact that the lensed CMB can be considered as a function of two
fields, the unlensed sky and the lensing potential, which in first
approximation can be assumed to be Gaussian (unmodified-LP case), even
if the non-linear evolution of large-scale structures produces
non-Gaussian features in the distribution of the projected potential
(MS-modified-LP case). In particular, the non-Gaussianity associated
with the large-scale structure induces non-Gaussian contributions to the distribution 
of the lensing potential such that its $n$-point correlator in Fourier space will
be non-vanishing for some value of $n$, and this, in turn, has an impact on
the connected part of the $n$-point correlators of the lensed CMB (see e.g. \cite{cumulants}).
Anyway, there exist also other phenomena that can produce
non-Gaussian effects in the lensed CMB, for instance the correlations between CMB lensing and the 
Sunyaev-Zeldovich effect.

Here, as a first characterization of the non-Gaussianity strength
produced by CMB lensing, we consider the PDFs of the \emph{difference}
$T$, $Q$, $U$ maps. In this way, we subtract the unlensed Gaussian sky
and isolate the non-Gaussian term which physically generates the
non-Gaussian signatures in the lensed CMB.  The obtained PDF in the
temperature case is showed in the lower panel of Fig.~\ref{TT_pdf}
(the $Q$ and $U$ cases have a similar trend).

The PDFs of the difference maps are characterized by a kurtosis excess
with respect to Gaussian distributions randomly generated with the
same mean and standard deviation values.  More precisely, averaging
over different realizations, the kurtosis excess is $\sim 2.41$ for
the difference $T$-map, $\sim 2.14$ for the difference $Q$-map and
$\sim 2.16$ for the difference $U$-map.  Actually we would expect that
the excess of non-linearities in the dark-matter distribution of the
Millennium Simulation should show up as an excess of non-Gaussianity
in the MS-modified-LP lensed difference $T$, $Q$, $U$ maps, with
respect to the unmodified-LP case.  However, the one-point statistic
is probably unable to capture this effect, and Fig.~\ref{TT_pdf} shows
in fact that there is not a significant difference between the
unmodified-LP and MS-modified-LP cases.  In this respect, we believe
that a more suitable estimator should be developed in order to detect
the contribution from the non-linear structure evolution to the total
non-Gaussian statistics of the lensed maps. We reserve this analysis
for future work.

\section{Conclusions}

We have constructed the first all-sky CMB temperature and polarization
lensed maps based on a high-resolution cosmological $N$-body
simulation, the Millennium Simulation.

To this purpose we have exploited the lensing potential map obtained
using the map-making procedure developed in \cite{Carbone_etal2008}
which integrates along the line-of-sight the MS dark matter structures
by stacking and randomizing the simulation boxes up to $z=127$.
Specifically, we have modified the LensPix code \citep{Lewis05} by
supplying it with the spherical harmonic coefficients extracted from
the MS lensing potential map and by implementing directly in the code
itself the large-scale structure adding technique
\citep[see][]{Carbone_etal2008} which allows to reinstate the
large-scale power in the angular lensing potential that is not
correctly sampled by the $N$-body simulation.  In this way we also
preserve the correct correlation between the lensing potential and the
ISW effect on multipoles $l \leq 400$ in the simulated temperature
anisotropies.

Using our modified version of the LensPix code, we have constructed
lensed CMB simulated maps with $\sim 5$ million pixels and an angular
resolution of $\sim 1.72'$, based on potential fields calculated on
$2560^3$ mesh cells from the Millennium Simulation.

After subtraction of the unlensed maps, the corresponding lensed $T$,
$Q$, $U$ simulated maps reflect clearly the same large scale structure
which is present in the modulus of the angular gradient of the lensing
potential map (see Fig.~\ref{TP_diff_maps}).

We have also constructed the maps of the scalar and pseudo-scalar
potentials $\psi_E$ and $\psi_B$ which are directly related to the
electric and magnetic types of polarization, respectively.  Their
difference maps, shown in
Figs.~\ref{psiE_diff_maps}-\ref{psiB_diff_maps}, present distinct
degree-scale distributions in the MS-modified-LP and unmodified-LP
cases (where the latter is obtained from the unmodified LensPix code
as previously explained), owing to the power transfer from non-linear
to large scales in the lensed $E$ and $B$ fields.

As a quantitative study of the simulated maps, we have performed power
spectrum and one-point statistics analyses. We find that the lensed
$TT$, $TE$, $EE$ and $BB$ power spectra, obtained using the MS dark
matter distribution, mostly overlap with the corresponding
semi-analytic expectations on a range of multipoles up to $l\sim 2500$
(see Figs.~\ref{TT_PS}-\ref{BB_PS}).  This latter result points out
the effectiveness of the Millennium Simulation in reproducing
correctly the findings of the theoretical approach.  Furthermore, an
excess of power is observable in the MS-modified-LP case on larger
multipoles, in particular for the $BB$ spectrum. We believe that this
excess originates from the accurate inclusion of non-linear power in
the Millennium Simulation, which is present in the MS lensing
potential power spectrum as well (Fig.~\ref{projpot_PS}).  This
outcome should be taken into account in the various delensing
approaches, since these non-linear effects can have an impact on the
quality of the reconstructed unlensed sky, particularly in view of the
detection of primary $B$-modes.

Finally, we have derived one-point statistics both of the simulated
$T$, $Q$, $U$ lensed maps and of the corresponding difference maps, in
the MS-modified-LP and unmodified-LP cases.  The comparison of the
resulting PDFs with respect to Gaussian random distributions with the
same mean and standard deviation does not show any statistical
difference when the total lensed maps are considered. On the other
hand, we find that the difference maps are characterized by a kurtosis
excess.  This result represents the distinctive weak-lensing effect of
inducing non-Gaussianity in the unlensed Gaussian CMB field.  This is
simply explained if in first approximation we think of the lensed
field as the product of two Gaussian fields, i.e. the lensing
potential and the primary unlensed CMB. Actually, correctly speaking,
the lensing potential map, derived by integration of the MS dark
matter distribution, preserves an intrinsic degree of non-Gaussianity
due to the non-linear evolution of the cosmic structures, which at
some level contributes to the non-Gaussian statistics of the simulated
lensed CMB.  Unfortunately the one-point statistic is not sufficient
to disentangle the two contributions, i.e. the dominant effect coming
from the product of two different fields, and the subdominant effect
coming from the non-linear matter evolution. This is the reason why in
Fig.~\ref{TT_pdf} there seems to be no difference between the
MS-modified-LP and unmodified-LP cases.  We reserve a more accurate
characterization of non-Gaussianity in CMB lensing statistics to a
future work.

\section*{Acknowledgments}
CC warmly thank L. Verde for precious suggestions, and E. Gaztanaga,
P. Fosalba, S. Leach, M. Liguori for helpful discussions.  CB thanks
B. Menard for helpful suggestions.  Some of the results in this paper
have been derived using the Hierarchical Equal Area Latitude
Pixelization of the sphere \citep{Healpix}.  CC is supported through a
Beatriu de Pinos grant.

\end{document}